\documentclass[10pt,twocolumn,letterpaper]{article}
\usepackage[accsupp]{axessibility}  
\usepackage{iccv}
\usepackage{times}
\usepackage{epsfig}
\usepackage{graphicx}
\usepackage{amsmath}
\usepackage{amssymb}

\usepackage{multirow}
\usepackage{booktabs}

\usepackage{colortbl}
\definecolor{Gray}{gray}{0.85}
\definecolor{blush}{rgb}{0.87, 0.36, 0.51}
\newcolumntype{a}{>{\columncolor{Gray}}c}
\newcommand\best[1]{\textbf{\color{blush}#1}}
\newcommand\bestclean[1]{\textbf{#1}}

\usepackage[pagebackref=true,breaklinks=true,letterpaper=true,colorlinks,bookmarks=false]{hyperref}

\iccvfinalcopy 
\def\iccvPaperID{6912} 

\ificcvfinal\fi
\begin{document}

\title{AdaptGuard: Defending Against Universal Attacks for Model Adaptation}

\author{Lijun Sheng$^{1,2}$, Jian Liang$^{2,3}$\thanks{To whom correspondence should be addressed.}\;, Ran He$^{2,3}$, Zilei Wang$^{1}$, Tieniu Tan$^{2,4}$ \\
$^{1}$ University of Science and Technology of China \\
$^{2}$ CRIPAC \& MAIS, Institute of Automation, Chinese Academy of Sciences \\
$^{3}$ University of Chinese Academy of Sciences
$^{4}$ Nanjing University \\
{\tt\small slj0728@mail.ustc.edu.cn, liangjian92@gmail.com}}

\maketitle
\ificcvfinal\thispagestyle{empty}\fi

\begin{abstract}
Model adaptation aims at solving the domain transfer problem under the constraint of only accessing the pretrained source models.
With the increasing considerations of data privacy and transmission efficiency, this paradigm has been gaining recent popularity.
This paper studies the vulnerability to universal attacks transferred from the source domain during model adaptation algorithms due to the existence of malicious providers.
We explore both universal adversarial perturbations and backdoor attacks as loopholes on the source side and discover that they still survive in the target models after adaptation.
To address this issue, we propose a model preprocessing framework, named AdaptGuard, to improve the security of model adaptation algorithms. 
AdaptGuard avoids direct use of the risky source parameters through knowledge distillation and utilizes the pseudo adversarial samples under adjusted radius to enhance the robustness.
AdaptGuard is a plug-and-play module that requires neither robust pretrained models nor any changes for the following model adaptation algorithms.
Extensive results on three commonly used datasets and two popular adaptation methods validate that AdaptGuard can effectively defend against universal attacks and maintain clean accuracy in the target domain simultaneously.
We hope this research will shed light on the safety and robustness of transfer learning.
Code is available at \url{https://github.com/TomSheng21/AdaptGuard}.
\end{abstract}

\begin{figure}[t]
    \centering
    \includegraphics[width=0.96\linewidth]{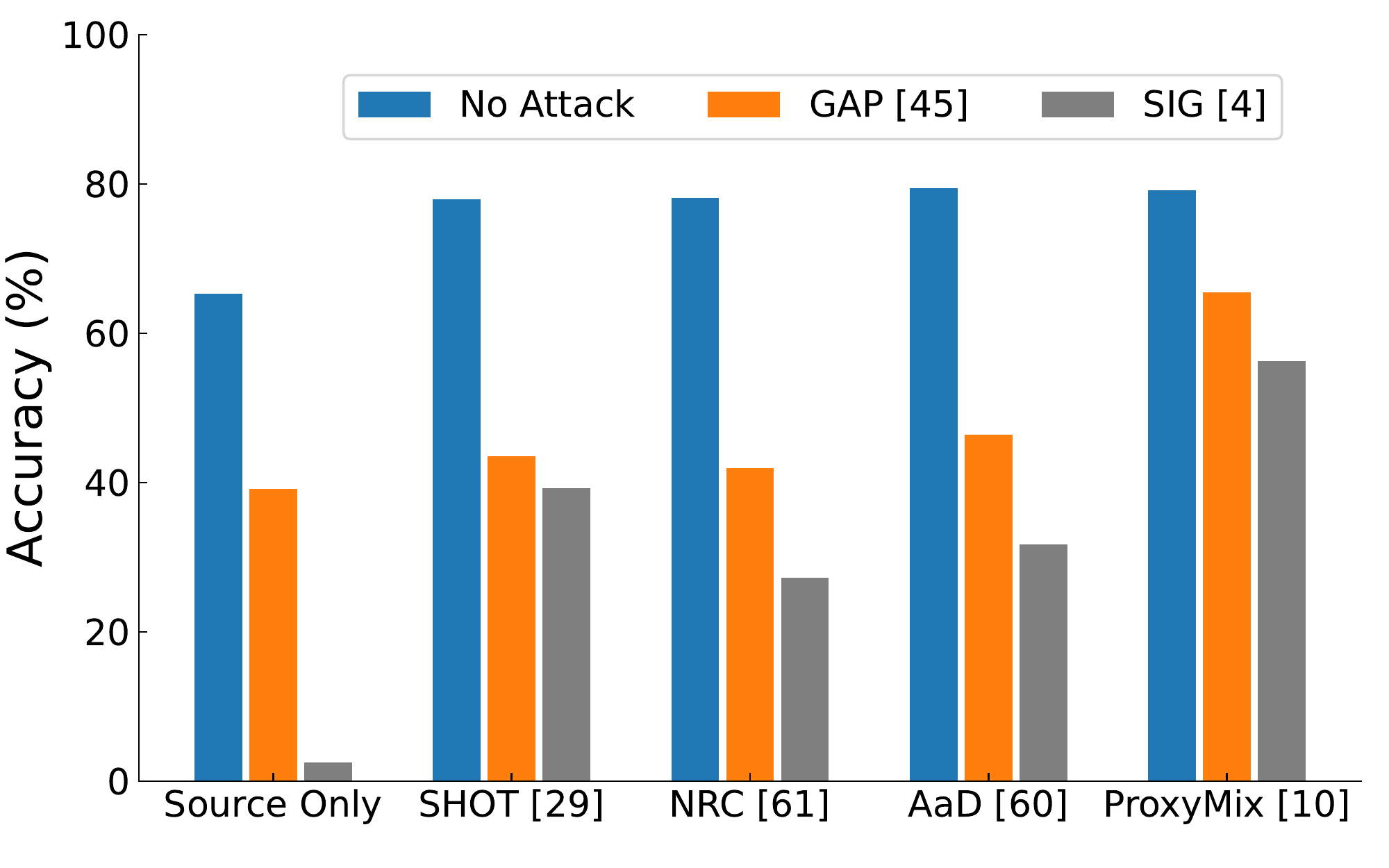}
    \caption{Target performance of five model adaptation methods under universal adversarial perturbation (\ie, GAP), and backdoor attack (\ie, SIG) on the task \textbf{A$\to$P} from \textbf{OfficeHome} \cite{venkateswara2017deep} dataset.}
    \label{fig:intro}
\end{figure}

\section{Introduction}
Over the last decade, great efforts have been dedicated to deep neural networks (DNN) \cite{krizhevsky2012imagenet, he2016deep}, which have made significant progress in computer vision, natural language processing, and many other applications.
Since the data distribution during deployment often differs from the training set, unsupervised domain adaptation \cite{ben2010theory, ganin2016domain, long2018conditional, liang2021domain, li2021probabilistic} is widely studied to improve the model performance across different domains and save the cost of extra annotations.
Growing attention is devoted to a new situation, model adaptation \cite{liang2020we, li2020model}, which restricts algorithms to only access the pretrained model from the source domain and the unlabeled target data.
This new paradigm \cite{liang2023comprehensive, yang2021exploiting, liang2021source, kurmi2021domain, ahmed2021unsupervised, wang2020tent, ding2022proxymix, qiu2021source} quickly becomes popular among the transfer learning field thanks to the appealing privacy-preserving property and the competitive performance to source-data-dependent methods.
Nowadays, the model adaptation scheme has been widely applied to various tasks, \eg, image classification \cite{liang2020we, yang2021exploiting, liang2021source, huang2021model}, semantic segmentation \cite{liu2021source, kundu2021generalize, you2021domain}, object detection \cite{xiong2021source, li2022source}.

As model adaptation methods always trust the source providers unconditionally \cite{liang2020we}, users may ignore the potential risks inside the source models.
In fact, providers would use inherent properties of models and the transferability of loopholes to attack clients' models which have been re-trained by adaptation algorithms.
This vulnerability may cause great security breaches when the models are deployed in real-world scenarios.
For example, if the street sign recognition system is under attack, it will make wrong decisions during driving.
And attacking the digital recognition network will cause great losses, especially in the financial systems and security fields.
The widely-used robustness metrics \cite{goodfellow2014explaining, madry2017towards} are based on image-specific adversarial perturbations.
As image-specific perturbations need multiple queries to the victim model per image, for the sake of simplicity and practicality, we choose to investigate image-agnostic perturbations, aka universal perturbations.
In our framework, universal perturbations are generated from the source side and directly used to execute attacks on the model after adaptation in the target domain.
This framework is more suitable and flexible and can be extended to many situations of different attacks.

This paper considers two kinds of universal perturbations from the source domain, \ie, universal adversarial perturbation (UAP) and backdoor attack.
UAP \cite{moosavi2017universal, poursaeed2018generative} aims to generate a quasi-imperceptible image-agnostic perturbation, which is able to alter the predictions on a large percentage of test samples.
Backdoor attack \cite{gu2017badnets, wu2022backdoorbench, chen2017targeted, barni2019new} poisons the training dataset with a designed trigger to embed the backdoor in the model, and when the trigger appears during testing, the prediction will be disturbed following the preset mode.
We empirically test the transferability of the selected universal attacks \cite{poursaeed2018generative, barni2019new} from the source domain towards existing model adaptation methods \cite{liang2020we, ding2022proxymix, yang2021exploiting, yang2022attracting}, and results are shown in Fig.~\ref{fig:intro}.
It is obvious that the performance under perturbation is much lower than the normal value, indicating that the model adaptation method will indeed inherit the risk from the source side.

To defend against the well-transferred universal attacks for model adaptation, we propose a model preprocessing framework, named AdaptGuard.
In particular, we conjecture the source model parameters to be risky and avoid using them directly as initialization.
Thereby, we extract the effective information from the source model via knowledge distillation \cite{hinton2015distilling}.
To against local perturbation, we introduce adversarial examples by projected gradient descent (PGD) \cite{madry2017towards} into distillation.
Further, we develop a radius-adjusting strategy that constrains adversarial examples to gradually become stronger, mitigating the negative impact on student network training at an early stage.
In this manner, the student model is expected to provide reliable initialization for the following model adaptation methods.
In the experiment, we validate the effectiveness of AdaptGuard for two popular model adaptation methods (\ie, SHOT \cite{liang2020we} and NRC \cite{yang2021exploiting}) on three datasets, including Office, OfficeHome, and DomainNet126. 
Besides, we provide the result under image-specific attack to show the flexibility of our framework.
Our contributions are summarized as follows:
\begin{itemize}
\item We examine the vulnerability of model adaptation against well-transferred universal attacks from malicious source providers, providing a new but critical perspective to existing methods.
\item We propose a simple yet effective framework, named AdaptGuard, to enhance the defense ability of model adaptation methods against universal attacks.
\item Extensive empirical results show that AdaptGuard performs effective defense and meanwhile maintains the domain transfer capacity.
\end{itemize}

\section{Related Work}

\subsection{Model Adaptation}
Model adaptation \cite{liang2023comprehensive, liang2020we, huang2021model}, also known as source-free unsupervised domain adaptation \cite{yang2021exploiting, qiu2021source, qiu2021source, ding2022proxymix, yang2022attracting, ding2023maps}, aims to transfer knowledge from the well-trained source model to the unlabeled target domain. 
Self-supervised learning-based methods \cite{ding2022proxymix, xia2021adaptive} are popular in the model adaptation task.
SHOT \cite{liang2020we}, as a representative work, provides an effective framework to align the target domain to the source hypothesis with information maximization and self-training strategy.
NRC \cite{yang2021exploiting} exploits the target distribution structure and encourages strong consistency between the adjacent samples for better alignment.
SHOT++ \cite{liang2021source} employs MixMatch \cite{berthelot2019mixmatch} to alleviate the noise caused by less-confident samples and uses a rotation prediction task to improve domain transfer.

Another mainstream idea to solve the model adaptation problem is source domain generation \cite{li2020model, kurmi2021domain, qiu2021source}.
SDDA \cite{kurmi2021domain} uses the conditional GAN \cite{mirza2014conditional} under the supervision of the source model to generate samples to replace absent source data.
CPGA \cite{qiu2021source} generates source avatar prototypes via contrastive learning and achieves adaptation with target pseudo labels.
SFIT \cite{hou2021visualizing} designs a two-branch framework to achieve image translation and fine-tunes the target model with the generated images.
However, generation-based methods always require additional models and are difficult to train.

\subsection{Universal Adversarial Perturbation}
Previous works \cite{szegedy2013intriguing, goodfellow2014explaining} observe deep neural networks are vulnerable to imperceptible perturbations, which can change the prediction of the networks.
Researchers \cite{moosavi2017universal} find that an image-agnostic perturbation is able to fool most samples during inference, which is termed as universal adversarial perturbation \cite{zhang2021survey}.
UAP \cite{moosavi2017universal} generates the perturbation by employing DeepFool \cite{moosavi2016deepfool} per image iteratively until the fooling rate reaches the desired value.
Some methods \cite{mopuri2018nag, poursaeed2018generative} use the generative ability of GANs \cite{goodfellow2014generative} to avoid directly optimizing the perturbation.
For short training time and simplicity of the framework, the following works \cite{zhang2020understanding, shafahi2020universal} improve the efficiency of searching universal perturbations without using generative models.

Corresponding to attack methods, various defense strategies \cite{moosavi2017universal, mummadi2019defending, shafahi2020universal, madry2017towards, ren2022towards} against UAP have also been explored.
Fine-tuning with perturbed images \cite{moosavi2017universal} is employed and proposes a shared adversarial training strategy \cite{mummadi2019defending} is designed to handle the trade-off between accuracy on clean examples and robustness against UAP.
Shafahi et al. \cite{shafahi2020universal} use universal adversarial training by optimizing a min-max problem using the alternating or simultaneous stochastic gradient.
There are also some works \cite{agarwal2022unsupervised, awais2021adversarial} study about robust domain adaptation, but they focus on the image-specific attack and always need additional information (\ie, robust ImageNet pretrained model).
Different from them, we consider UAP which is more suitable for model adaptation and design defense strategy with external resources.

\begin{figure}[t]
		\centering
		\small
		\setlength\tabcolsep{1mm}
		\renewcommand\arraystretch{0.1}
		\begin{tabular}{ccc}
			\includegraphics[width=0.3\linewidth, clip]{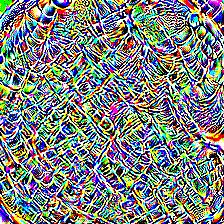} &
			\includegraphics[width=0.3\linewidth, clip]{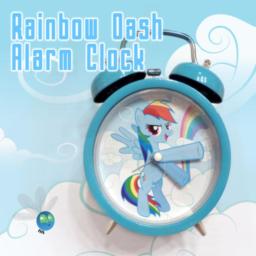} & 
			\includegraphics[width=0.3\linewidth, clip]{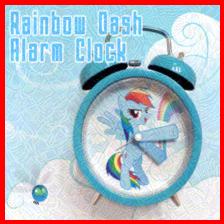} \\
            \\[1mm]
            (a) Visualize UAP. & (b) Original `clock'. & (c) + UAP 
            \cite{moosavi2017universal}.  \\
            \\[1mm]
			\includegraphics[width=0.3\linewidth, clip]{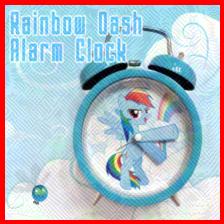}&
            \includegraphics[width=0.3\linewidth, clip]{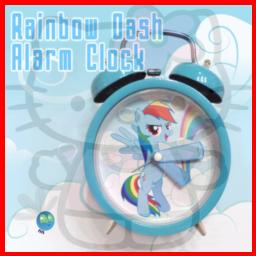}&
            \includegraphics[width=0.3\linewidth, clip]{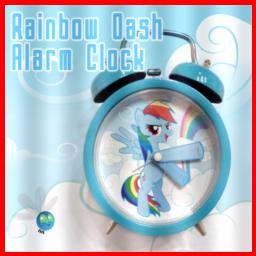}\\
            \\[1mm]
            (d) + GAP \cite{poursaeed2018generative}. & (e) + Blended \cite{chen2017targeted}. & (f) + SIG \cite{barni2019new}. \\
            \\[1mm]
		\end{tabular}
		\caption{Visualization of perturbations and examples of an artistic image `clock' from \textbf{OfficeHome} \cite{venkateswara2017deep} under four attacks. The red border is only used to distinguish the attacked images here.}
		\label{fig:attack}
		\vspace{-10pt}
\end{figure} 

\subsection{Backdoor Attack}
Backdoor attacks \cite{gu2017badnets, wu2022backdoorbench, li2022backdoor, zhang2023red} poison the training dataset with a well-designed trigger to embed a backdoor into the model.
The victim model behaves normally on clean samples, but when the trigger appears, it makes predictions expected by attackers.
BadNets \cite{gu2017badnets} first proposes the backdoor attack by putting a pattern of bright pixels in the corner of the MNIST images.
Blended \cite{chen2017targeted} achieves backdoor poisoning attacks with a fixed image trigger through a blended injection strategy.
The following poison methods \cite{barni2019new, shafahi2018poison, zeng2021rethinking} design invisible and natural triggers with effective poisoning manners.
Besides using a single pattern as the trigger, there are also backdoor attacks \cite{nguyen2020input, nguyen2021wanet} that control the training process and learn the trigger during training.

We focus on backdoor defense in the post-training stage \cite{wu2022backdoorbench, wu2021adversarial, wang2019neural, guan2022few} which mitigates the impact of the backdoor on a well-trained model.
NAD \cite{li2021neural} utilizes a teacher network to guide the fine-tuning of the victim model.
ANP \cite{wu2021adversarial} prunes sensitive neurons under adversarial neuron perturbations to remove the injected backdoor.
These defense methods always require the part of clean training data, so they can not be utilized in the model adaptation scenarios.

\section{Universal Attacks in Model Adaptation}

Most model adaptation methods follow the framework proposed in SHOT \cite{liang2020we} and train the source model $f_s$ using empirical risk minimization with label smoothing technique \cite{muller2019does} on labeled source data $\mathcal{D}_s = \{(x_i^s, y_i^s)\}_{i=1}^{N_s}$.
Then, users utilize the well-trained source model and unlabeled target data $\mathcal{D}_t = \{x_i^t\}_{i=1}^{N_t}$ to build the target model $f_t$ through their own adaptation algorithms.

Existing adaptation algorithms always ignore the risk from the source domain, whose training procedure is actually uncontrolled in real scenarios.
In this section, we discuss two types of universal attacks that are easy to implement by source domain providers.

\textbf{Remark.}
We focus on universal attacks from the source side in this section, which are also the main attack methods considered in our framework.
Different from most image-specific attacks, universal attacks \textit{need no queries to the victim model per image}, which are more convenient to calculate and implement.

\subsection{Universal Adversarial Perturbation}

Previous work \cite{moosavi2017universal} observes the existence of universal adversarial perturbation $v$ which misleads the classifier on almost test images.
The $\ell_p$ norm of the perturbation $v$ is less than a predefined value, guaranteeing its stealthiness and semantic invariance.
$v$ is calculated from the well-trained model and can be generalized well on the whole data.
The attacker obtains the perturbation through directly optimizing the perturbation \cite{moosavi2017universal} or training a generative network \cite{poursaeed2018generative} instead.

In model adaptation problems, model providers can calculate universal perturbations using the model $f_s$ and dataset $\mathcal{D}_s$ in the source domain.
The perturbation $v$ will satisfy the following requirements:
\begin{equation}
    \underset{x,y\sim\mathcal{D}_s}{\mathbb{P}} (f_s (x) \neq f_s (x+v)) \geq \delta , \
    s.t. \ \| v \|_p \leq \xi ,
    \label{eq:UAP in model adaptation}
\end{equation}
where $\mathbb{P}$ denotes probability calculation and $\delta<1.0$ represents the desired fooling rate.
The constraint means that the $\ell_p$ norm of $v$ is not greater than $\xi$.
As shown in Fig.~\ref{fig:attack} (c-d), the images after adding universal perturbations are quasi-imperceptible to humans.
UAP is a powerful attack on current model adaption techniques due to its great stealthiness and natural generation process.

\begin{figure*}[t]
    \centering
    \includegraphics[width=0.9\linewidth]{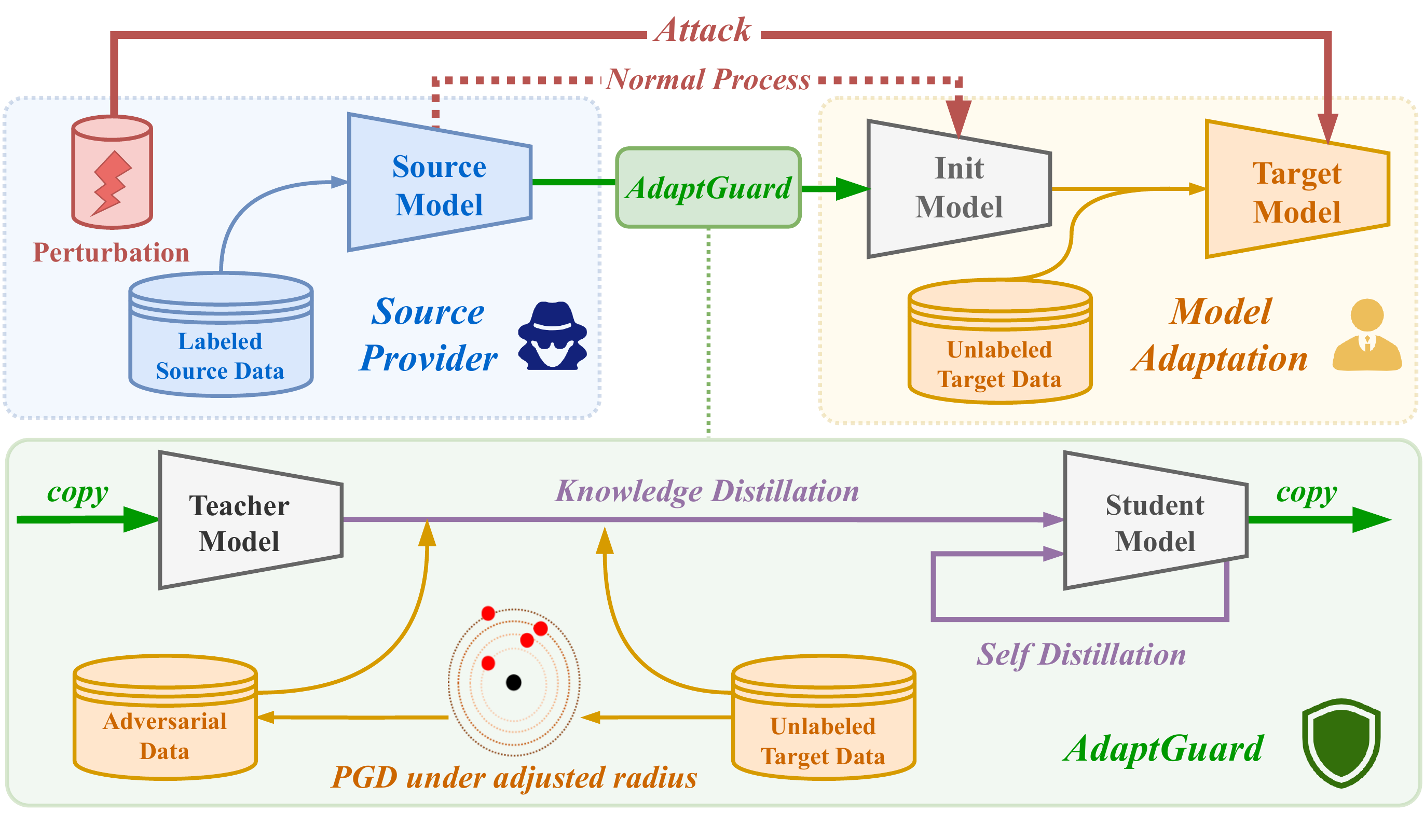}
    \caption{An overview of the proposed AdaptGuard framework. 
Knowledge distillation strategy is used to extract information about unlabeled data from source models. Adversarial examples under adjusted radius are introduced during distillation to further improve the model defense ability. Finally, the student network can be regarded as a reliable initialization for downstream model adaptation methods.}
    \label{fig:framework}
\end{figure*}

\subsection{Backdoor Attack}
Backdoor attack methods poison part of the training set with the designed trigger and the predefined label to embed backdoors into networks.

In the domain adaptation task, malicious model providers randomly select a portion of the source data to poison and specify a targeted category label $y^b$.
Blended \cite{chen2017targeted} takes a fixed image as a trigger that has the same size as the training data $x_i^{s}$.
The backdoor source data $x_i^{b,s}$ is calculated through a mixup manner as follows,
\begin{equation}
   \textbf{Blended}: x_i^{b,s} = (1-\alpha) \cdot x_i^{s} + \alpha \cdot {I}_{tri},
    \label{eq:blended in model adaptation}
\end{equation}
where ${I}_{tri}$ denotes the designed trigger and $\alpha$ controls the mix ratio.
Another pattern of trigger \cite{barni2019new} uses a designed perturbation that can be directly added to the training data.
This perturbation needs to have a bounded amplitude and the backdoor source data must be clamped to the meaningful interval as,
\begin{equation}
    \textbf{SIG}: x_i^{b,s} = {clamp}( x_i^{s} + {\Delta}_{tri}, 0, 255),
    \label{eq:sig in model adaptation}
\end{equation}
where ${\Delta}_{tri}$ represents the trigger perturbation and $clamp$ denotes the clamp operation to make sure all pixel values are in the range between 0 and 255.

The providers replace the selected set with its poisoned version $\{(x_i^{b,s}, y^b)\}_{i=1}^{N_b}$ and train the backdoored source model $f_s^b$.
Fig.~\ref{fig:attack} (e-f) shows the attacked images under the above two backdoor attacks.
As the victim model maintains normal performance on clean samples, backdoor attacks are difficult to detect.
Even if users re-train the victim model with the target data, the backdoor can still exist and result in a successful attack sometimes.

\setlength{\tabcolsep}{4.0pt}
    \begin{table*}[!t]
        \centering
        \caption{Accuracies (\%) of various defense methods against four attacks on \textbf{Office} \cite{saenko2010adapting} dataset for model adaptation (ResNet-50). Best accuracy of clean samples (\bestclean{bold}), best accuracy of attacked samples (\best{bold red}).}
        \vspace{3mm}
        \resizebox{0.9\textwidth}{!}{
            \begin{tabular}{cl|ccca|ccca|ccca|ccca}
            \toprule
            \multicolumn{2}{c|}{Attack} & \multicolumn{4}{c|}{UAP \cite{moosavi2017universal}} & \multicolumn{4}{c|}{GAP \cite{poursaeed2018generative}} & \multicolumn{4}{c|}{Blended \cite{chen2017targeted}} & \multicolumn{4}{c}{SIG \cite{barni2019new}} \\
            \midrule
            \multicolumn{2}{c|}{Task} & A$\to$ & D$\to$ & W$\to$ & Avg & A$\to$ & D$\to$ & W$\to$ & Avg & A$\to$ & D$\to$ & W$\to$ & Avg & A$\to$ & D$\to$ & W$\to$ & Avg \\ 
            \midrule
            \multicolumn{2}{c|}{Clean (Source Only)} & 78.8 & 78.4 & 80.1 & 79.1 & 78.8 & 78.4 & 80.1 & 79.1 & 76.6 & 77.0 & 79.9 & 77.9 & 77.3 & 77.1 & 79.2 & 77.8 \\
            \multicolumn{2}{c|}{Attack (Source Only)} &  37.6 & 47.0 & 42.5 & 42.4 & 13.1 & 46.4 & 43.2 & 34.2 & 0.3 & 0.6 & 1.3 & 0.7 & 1.1 & 13.5 & 4.0 & 6.2 \\
            \midrule
            \midrule
            \multirow{4}{3.5em}{Clean (SHOT)} & \multicolumn{1}{|l|}{SHOT \cite{liang2020we}} & 92.6 & \bestclean{85.9} & \bestclean{88.0} & \bestclean{88.8} & 92.6 &  \bestclean{85.9} &  \bestclean{88.0} &  \bestclean{88.8} & \bestclean{92.7} & \bestclean{85.4} & \bestclean{86.9} & \bestclean{88.4} & \bestclean{93.7} & \bestclean{85.4} & 87.0 & \bestclean{88.7} \\
            & \multicolumn{1}{|l|}{+ANP \cite{wu2021adversarial}} & 90.7 & \bestclean{85.9} & 87.1 & 87.9 & 90.7 &  \bestclean{85.9} & 87.1 & 87.9 & 90.0 & 85.2 & 83.9 & 86.4 & 89.0 & 85.0 & \bestclean{87.2} & 87.0  \\
            & \multicolumn{1}{|l|}{+TRADES \cite{zhang2019theoretically}} & 91.0 & 85.7 & 86.7 & 87.8 & 91.0 & 85.7 & 86.7 & 87.8 & 91.4 & 84.9 & 85.6 & 87.3 & 91.9 & 84.7 & 84.9 & 87.2 \\
            & \multicolumn{1}{|l|}{+PGD \cite{madry2017towards}} & \bestclean{92.8} & 85.4 & 86.3 & 88.2 &  \bestclean{92.8} & 85.4 & 86.3 & 88.2 & 92.5 & 84.6 & 85.4 & 87.5 & 93.6 & 84.6 & 85.5 & 87.9 \\
            & \multicolumn{1}{|l|}{+AdaptGuard} & 88.2 & 83.7 & 85.4 & 85.8 & 88.2 & 83.7 & 85.4 & 85.8 & 88.4 & 83.7 & 85.2 & 85.8 & 88.1 & 83.0 & 84.6 & 85.2  \\
            \midrule
            \multirow{4}{3em}{Attack (SHOT)} & \multicolumn{1}{|l|}{SHOT \cite{liang2020we}} & 70.6 & 72.0 & 64.9 & 69.2 & 40.5 & 68.0 & 64.2 & 57.6 & 2.9 & 27.7 & 19.9 & 16.8 & 15.2 & 60.6 & 33.5 & 36.4 \\
            & \multicolumn{1}{|l|}{+ANP \cite{wu2021adversarial}} & 71.6 & 71.5 & 67.4 & 70.2 & 44.1 & 64.7 & 64.6 & 57.8 & 6.5 & 36.5 & 11.3 & 18.1 & 14.8 & 60.5 & 36.1 & 37.1 \\
            & \multicolumn{1}{|l|}{+TRADES \cite{zhang2019theoretically}} & 85.4 & \best{85.1} & 81.2 & 83.9 & \best{67.0} & \best{83.6} & 76.4 & 75.6 & 23.5 & 50.3 & 31.2 & 35.0 & 38.9 & 75.0 & 47.8 & 53.9 \\
            & \multicolumn{1}{|l|}{+PGD \cite{madry2017towards}} & \best{87.6} & 84.7 & 81.9 & 84.7 & 65.5 & 83.0 & 79.1 & 75.8 & 25.5 & 49.2 & 36.7 & 37.1 & 50.8 & \best{77.8} & 52.8 & \best{60.5}  \\
            & \multicolumn{1}{|l|}{+AdaptGuard} & 86.4 & 83.7 & \best{85.1} & \best{85.1} & 61.8 & \best{83.6} & \best{82.8} & \best{76.1} & \best{60.3} & \best{63.1} & \best{61.0} & \best{61.5} & \best{59.7} & 61.4 & \best{60.4} & \best{60.5} \\
            \midrule
            \midrule
            \multirow{4}{3em}{Clean (NRC)} & \multicolumn{1}{|l|}{NRC \cite{yang2021exploiting}} & 90.9 & \bestclean{86.2} & \bestclean{87.1} & 88.1 & 90.9 & \bestclean{86.2} & \bestclean{87.1} & 88.1 & 91.3 & \bestclean{86.8} & \bestclean{86.8} & \bestclean{88.3} & 90.1 & \bestclean{86.7} & 86.9 & \bestclean{87.9}  \\ 
            & \multicolumn{1}{|l|}{+ANP \cite{wu2021adversarial}} &  91.5 & \bestclean{86.2} & 86.9 & \bestclean{88.2} & 91.5 & \bestclean{86.2} & 86.9 & \bestclean{88.2} & 91.1 & 86.2 & 86.3 & 87.9 & 89.8 & 86.5 & \bestclean{87.3} & \bestclean{87.9}  \\
            & \multicolumn{1}{|l|}{+TRADES \cite{zhang2019theoretically}} & 89.3 & 84.9 & 84.0 & 86.1 & 89.3 & 84.9 & 84.0 & 86.1 & 89.5 & 84.9 & 83.7 & 86.1 & 87.7 & 83.7 & 82.6 & 84.7 \\ 
            & \multicolumn{1}{|l|}{+PGD \cite{madry2017towards}} & \bestclean{92.6} & 82.8 & 84.2 & 86.5 & \bestclean{92.6} & 82.8 & 84.2 & 86.5 & \bestclean{92.2} & 82.4 & 84.5 & 86.4 & \bestclean{92.0} & 81.5 & 84.2 & 85.9  \\ 
            & \multicolumn{1}{|l|}{+AdaptGuard} & 91.0 & 85.3 & 85.2 & 87.2 & 91.0 & 85.3 & 85.2 & 87.2 & 91.0 & 84.9 & 85.8 & 87.2 & 89.6 & 82.7 & 85.3 & 85.9  \\ 
            \midrule
            \multirow{4}{3em}{Attack (NRC)} & \multicolumn{1}{|l|}{NRC \cite{yang2021exploiting}} & 62.0 & 63.8 & 59.9 & 61.9 & 35.6 & 63.4 & 60.5 & 53.2 & 1.3 & 19.4 & 14.9 & 11.8 & 9.4 & 41.8 & 20.1 & 23.8 \\ 
            & \multicolumn{1}{|l|}{+ANP \cite{wu2021adversarial}} & 64.4 & 64.2 & 60.5 & 63.0 & 39.3 & 59.1 & 59.6 & 52.6 & 2.9 & 23.6 & 10.3 & 12.2 & 12.8 & 46.7 & 26.2 & 28.5  \\ 
            & \multicolumn{1}{|l|}{+TRADES \cite{zhang2019theoretically}} & 86.9 & \best{84.1} & 80.1 & 83.7 & \best{74.8} & 82.9 & 75.1 & 77.6 & 28.2 & 46.9 & 31.9 & 35.6 & 43.0 & 73.1 & 46.6 & 54.3  \\ 
            & \multicolumn{1}{|l|}{+PGD \cite{madry2017towards}} & \best{90.1} & 82.7 & 83.1 & \best{85.3} & 74.5 & 81.5 & 79.8 & \best{78.6} & 31.5 & 52.3 & 42.6 & 42.1 & 58.0 & 76.2 & 62.1 & 65.4  \\ 
            & \multicolumn{1}{|l|}{+AdaptGuard} & 86.4 & 83.7 & \best{85.1} & 85.1 & 63.6 & \best{84.1} & \best{80.5} & 76.1 & \best{58.7} & \best{61.3} & \best{59.5} & \best{59.8} & \best{84.2} & \best{78.7} & \best{81.7} & \best{81.5}  \\
            \bottomrule
            \end{tabular}
        }
    \label{tab: result office}
    \end{table*}

\section{Defending Against Universal Attacks}
As common image-agnostic attack methods, universal adversarial perturbations and backdoor attacks are still threats to the target model after model adaptation.
We proposed a model preprocessing defense framework termed AdaptGuard, to defend against potential attacks from the source provider.

\subsection{Knowledge Distillation from the Source Model}
To avoid directly using the risky source model as initialization during adaptation, we choose to extract information correlative with the target domain via knowledge distillation (KD) \cite{hinton2015distilling}.
Let $f$ be the network and $f(x)$ be the logit of input $x$ given by $f$. 
The source model $f_s$ is regarded as the teacher network. 
We init the student network $f_{kd}$ with ImageNet \cite{deng2009imagenet} pretrained parameters and init the last fully-connected layer randomly.
KD optimizes the student network by minimizing the discrepancy between the predictions from the two networks.
In the target domain $\mathcal{D}_t = \{x_i^t\}_{i=1}^{N_t}$, we take the KL divergence as the discrepancy criterion and the optimization process can be expressed as follows:
\begin{equation}
	\mathcal{L}_{kd}(f_{kd}) =
	\mathbb{E}_{x_t\in \mathcal{D}_t} \mathcal{KL}\left(\sigma(f_s(x_t)) \ \| \ \sigma(f_{kd}(x_t))\right),
	\label{eq:method_kd}
\end{equation}
where $\mathcal{KL}(\cdot\|\cdot)$ denotes the Kullback–Leibler divergence of two probability distributions and $\sigma$ is the softmax operation.

We follow \cite{liang2022dine} and use a self-distillation strategy to mitigate the effects of noise in the teacher network.
This paradigm uses a dynamic supervised signal $P(x_t)$ init with the teacher network prediction $f_s(x_t)$.
The student network $f_{kd}$ updates $P(x_t)$ with its prediction through exponential moving average (EMA) every few iterations.
The self-distillation strategy introduces the self-training pattern to the standard distillation process.
The final loss function is given by,
\begin{equation}
    \begin{aligned}
        \mathcal{L}_{kd} &=
		\mathbb{E}_{x_t\in \mathcal{D}_t}\  \mathcal{KL}\left(\sigma(P(x_t)) \ \| \ \sigma(f_{kd}(x_t))\right), \\
        P&(x_t) \leftarrow \gamma P(x_t) + (1-\gamma) f_{kd}(x_t),
    \end{aligned}
    \label{eq:kd_clean}
\end{equation}
where $\gamma$ is a hyperparameter representing the weight of past prediction in EMA.

\setlength{\tabcolsep}{3.0pt}
    \begin{table*}[!t]
        \centering
        \caption{Accuracies (\%) of various defense methods against four attacks on \textbf{OfficeHome} \cite{venkateswara2017deep} dataset for model adaptation (ResNet-50).}
        \vspace{3mm}
        \resizebox{1.0\textwidth}{!}{
            \begin{tabular}{ll|cccca|cccca|cccca|cccca}
            \toprule
            \multicolumn{2}{c|}{Attack} & \multicolumn{5}{c|}{UAP \cite{moosavi2017universal}} & \multicolumn{5}{c|}{GAP \cite{poursaeed2018generative}} & \multicolumn{5}{c|}{Blended \cite{chen2017targeted}} & \multicolumn{5}{c}{SIG \cite{barni2019new}} \\
            \midrule
            \multicolumn{2}{c|}{Task} & A$\to$ & C$\to$ & P$\to$ & R$\to$ & Avg & A$\to$ & C$\to$ & P$\to$ & R$\to$ & Avg & A$\to$ & C$\to$ & P$\to$ & R$\to$ & Avg & A$\to$ & C$\to$ & P$\to$ & R$\to$ & Avg \\ 
            \midrule
            \multicolumn{2}{c|}{Clean (Source Only)} & 60.7 & 59.3 & 55.0 & 62.4 & 59.4 & 60.7 & 59.3 & 55.0 & 62.4 & 59.4 & 60.5 & 56.0 & 54.0 & 60.9 & 57.9 & 59.7 & 56.4 & 53.6 & 60.7 & 57.6 \\
            \multicolumn{2}{c|}{Attack (Source Only)} & 29.7 & 13.8 & 12.4 & 15.6 & 17.9 & 37.0 & 16.1 & 16.2 & 22.0 & 22.8 & 2.8 & 1.0 & 0.9 & 1.1 & 1.4 & 3.4 & 4.2 & 3.6 & 1.3 & 3.1  \\
            \midrule
            \midrule
            \multirow{4}{3.5em}{Clean (SHOT)} & \multicolumn{1}{|l|}{SHOT \cite{liang2020we}} & \bestclean{71.9} & \bestclean{74.6} & \bestclean{68.6} & \bestclean{72.3} & \bestclean{71.9} & \bestclean{71.9} & \bestclean{74.6} & \bestclean{68.6} & \bestclean{72.3} & \bestclean{71.9} & \bestclean{71.5} & \bestclean{73.2} & \bestclean{67.6} & \bestclean{72.3} & \bestclean{71.1} & \bestclean{71.5} & \bestclean{73.7} & \bestclean{67.8} & \bestclean{71.7} & \bestclean{71.2}  \\
            & \multicolumn{1}{|l|}{+ANP \cite{wu2021adversarial}} & 70.0 & 72.8 & 66.9 & 71.6 & 70.3 & 70.0 & 72.8 & 66.9 & 71.6 & 70.3 & 69.7 & 66.5 & 65.7 & 2.0 & 51.0 & 69.1 & 65.1 & 66.8 & 2.4 & 50.8  \\
            & \multicolumn{1}{|l|}{+TRADES \cite{zhang2019theoretically}} & 70.6 & 73.0 & 67.5 & 71.2 & 70.6 & 70.6 & 73.0 & 67.5 & 71.2 & 70.6 & 70.3 & 72.3 & 67.1 & 71.0 & 70.2 & 70.3 & 72.8 & 67.1 & 70.4 & 70.2  \\
            & \multicolumn{1}{|l|}{+PGD \cite{madry2017towards}} & 68.9 & 72.0 & 66.1 & 70.1 & 69.3 & 68.9 & 72.0 & 66.1 & 70.1 & 69.3 & 68.7 & 70.7 & 66.1 & 70.3 & 68.9 & 68.6 & 71.3 & 66.0 & 69.9 & 68.9  \\
            & \multicolumn{1}{|l|}{+AdaptGuard} & 70.2 & 70.9 & 65.0 & 68.3 & 68.6 & 70.2 & 70.9 & 65.0 & 68.3 & 68.6 & 70.1 & 69.4 & 64.4 & 68.6 & 68.1 & 69.6 & 70.0 & 64.1 & 68.1 & 68.0  \\
            \midrule
            \multirow{4}{3em}{Attack (SHOT)} & \multicolumn{1}{|l|}{SHOT \cite{liang2020we}} & 50.6 & 52.9 & 40.4 & 35.7 & 44.9 & 47.6 & 44.2 & 38.1 & 32.0 & 40.5 & 37.2 & 31.6 & 27.7 & 25.7 & 30.6 & 32.3 & 36.6 & 35.1 & 25.4 & 32.3  \\
            & \multicolumn{1}{|l|}{+ANP \cite{wu2021adversarial}} & 55.8 & 55.8 & 41.9 & 38.3 & 48.0 & 52.9 & 47.9 & 37.7 & 38.0 & 44.1 & 37.9 & 38.2 & 29.8 & 1.1 & 26.7 & 50.9 & 47.0 & 45.7 & 1.3 & 36.2  \\
            & \multicolumn{1}{|l|}{+TRADES \cite{zhang2019theoretically}} & \best{69.8} & 70.8 & \best{65.1} & 66.5 & 68.1 & \best{68.8} & 67.5 & 61.7 & 64.0 & 65.5 & 48.3 & 46.4 & 37.8 & 40.6 & 43.3 & 56.6 & 56.5 & 55.7 & 53.8 & 55.6  \\
            & \multicolumn{1}{|l|}{+PGD \cite{madry2017towards}} & 68.8 & \best{71.2} & 65.0 & \best{68.2} & \best{68.3} & 68.6 & \best{70.2} & \best{62.9} & \best{65.9} & \best{66.9} & 52.6 & 52.1 & 46.1 & 48.2 & 49.7 & 62.1 & 64.8 & \best{61.3} & 60.2 & 62.1  \\
            & \multicolumn{1}{|l|}{+AdaptGuard} & 70.0 & 70.1 & 64.0 & 67.2 & 67.8 & 68.6 & 68.1 & 61.4 & 64.3 & 65.6 & \best{53.4} & \best{54.0} & \best{47.8} & \best{51.8} & \best{51.7} & \best{62.3} & \best{65.8} & 59.6 & \best{61.7} & \best{62.4} \\
            \midrule
            \midrule
            \multirow{4}{3em}{Clean (NRC)} & \multicolumn{1}{|l|}{NRC \cite{yang2021exploiting}} & \bestclean{72.4} & \bestclean{75.0} & \bestclean{67.8} & \bestclean{71.8} & \bestclean{71.8} & \bestclean{72.4} & \bestclean{75.0} & \bestclean{67.8} & \bestclean{71.8} & \bestclean{71.8} & \bestclean{71.6} & \bestclean{74.3} & \bestclean{67.6} & \bestclean{70.9} & \bestclean{71.1} & \bestclean{71.7} & \bestclean{74.1} & \bestclean{67.7} & \bestclean{71.0} & \bestclean{71.1} \\ 
            & \multicolumn{1}{|l|}{+ANP \cite{wu2021adversarial}} & 71.1 & 73.5 & 67.0 & 71.3 & 70.7 & 71.1 & 73.5 & 67.0 & 71.3 & 70.7 & 70.4 & 70.9 & 66.5 & 5.8 & 53.4 & 70.8 & 70.5 & 66.6 & 4.6 & 53.1 \\ 
            & \multicolumn{1}{|l|}{+TRADES \cite{zhang2019theoretically}} & 64.9 & 65.8 & 58.7 & 66.1 & 63.9 & 64.9 & 65.8 & 58.7 & 66.1 & 63.9 & 64.6 & 64.0 & 57.9 & 65.1 & 62.9 & 63.2 & 64.1 & 56.6 & 64.5 & 62.1  \\ 
            & \multicolumn{1}{|l|}{+PGD \cite{madry2017towards}} & 65.1 & 68.3 & 61.6 & 66.5 & 65.4 & 65.1 & 68.3 & 61.6 & 66.5 & 65.4 & 64.3 & 67.4 & 61.5 & 65.4 & 64.6 & 63.9 & 68.6 & 62.0 & 65.3 & 64.9 \\
            & \multicolumn{1}{|l|}{+AdaptGuard} & 70.9 & 70.8 & 64.5 & 68.1 & 68.6 & 70.9 & 70.8 & 64.5 & 68.1 & 68.6 & 70.6 & 68.3 & 63.6 & 67.5 & 67.5 & 69.8 & 68.5 & 63.0 & 67.7 & 67.2  \\ 
            \midrule
            \multirow{4}{3em}{Attack (NRC)} & \multicolumn{1}{|l|}{NRC \cite{yang2021exploiting}} & 45.6 & 47.4 & 33.2 & 29.2 & 38.8 & 44.8 & 40.2 & 33.8 & 31.5 & 37.6 & 25.1 & 32.7 & 22.2 & 20.2 & 25.0 & 21.2 & 31.7 & 24.9 & 19.1 & 24.2  \\ 
            & \multicolumn{1}{|l|}{+ANP \cite{wu2021adversarial}} & 51.4 & 52.0 & 37.2 & 31.1 & 42.9 & 49.6 & 44.0 & 35.2 & 35.7 & 41.1 & 28.8 & 36.4 & 24.6 & 2.7 & 23.1 & 38.5 & 38.6 & 27.1 & 3.4 & 26.9  \\ 
            & \multicolumn{1}{|l|}{+TRADES \cite{zhang2019theoretically}} & 61.8 & 58.7 & 50.5 & 55.9 & 56.7 & 59.6 & 52.8 & 44.9 & 53.3 & 52.6 & 34.4 & 30.4 & 22.1 & 27.9 & 28.7 & 34.3 & 32.9 & 29.3 & 28.8 & 31.3  \\ 
            & \multicolumn{1}{|l|}{+PGD \cite{madry2017towards}} & 64.1 & 65.3 & 58.0 & 62.4 & 62.5 & 63.4 & 62.1 & 54.3 & 59.4 & 59.8 & 47.4 & 43.2 & 33.9 & 37.7 & 40.5 & 50.7 & 54.8 & 46.2 & 44.9 & 49.2  \\ 
            & \multicolumn{1}{|l|}{+AdaptGuard} & \best{70.4} & \best{68.5} & \best{63.3} & \best{66.8} & \best{67.3} & \best{69.5} & \best{66.7} & \best{61.5} & \best{64.4} & \best{65.5} & \best{53.0} & \best{51.5} & \best{45.6} & \best{50.8} & \best{50.3} & \best{61.1} & \best{63.2} & \best{55.9} & \best{62.1} & \best{60.6} \\
            \bottomrule
            \end{tabular}
        }
    \label{tab: result office-home} 
    \end{table*}

\subsection{Adversarial Examples under Adjusted Radius}
Introducing adversarial examples into the training stage is an effective method to improve robustness.
After training with adversarial examples, the decision boundary is far away from the data point, so that the neighbor block of the sample in the input space will have the same prediction.
Based on this, we incorporate adversarial samples in the distillation process.
Adversarial examples $x_{adv}$ are calculated by projected gradient descent (PGD) \cite{madry2017towards} on the negative loss function in a multi-step manner:
\begin{equation}
    x_{t,adv}^{(i+1)} = \Pi_{x+\epsilon} \left( x_{t,adv}^{(i)} + \alpha \cdot sgn(\nabla_x L(x,y)) \right),
    \label{eq:PGD}
\end{equation}
where $\alpha$ represents the step size in each iteration, $i$ is the current iteration number and $\epsilon$ controls the size of the searching space to ensure the similarity to the original.
For the loss function $L$ in the formula, PGD \cite{madry2017towards} adopts the cross entropy loss which requires the ground truth of the input sample.
We choose to replace it with the pseudo label in the model adaptation task.
The pseudo label is also updated with the model during the distillation process.
The optimization objective of KD considering both the original sample and the adversarial sample is as follows:
\begin{equation}
    \begin{aligned}
		\mathcal{L}_{kd} =
		\mathbb{E}&_{x_t\in \mathcal{D}_t} \frac{1}{2} \mathcal{KL}\left(\sigma(P(x_t)) \ \| \ \sigma(f_{kd}(x_t))\right) + \\
	     &\frac{1}{2} \mathcal{KL}\left(\sigma(P(x_t)) \ \| \ \sigma(f_{kd}(x_{t,adv}))\right).
	\end{aligned}
	\label{eq:method_kd_all}
\end{equation}

Adversarial examples can help us defend against attacks, they also disturb the training process.
In the early stage of the distillation process, the student network does not perform well in the target domain.
In this period, strong adversarial examples will interfere with learning the target domain knowledge, resulting in performance degradation on the target clean samples.
To maintain network performance and not introduce extra hyperparameters, we employ a radius-adjusting strategy.
We restrict the radius $\epsilon$ of the adversarial example's searching space to grow linearly from 0 to the maximum value.
Adversarial examples also follow a change from weak to strong.
The initial adversarial examples are similar to the original and will have fewer side effects during the distillation process.
Later when the student model already has stable performance on the target domain, strong adversarial examples can be exploited to further improve the defense capability.

\textbf{Discussions of the overall defense process.}
We provide a summary of AdaptGuard here.
Users of model adaptation methods should not directly take the well-trained model from the source domain as the initialization of the follow-up algorithm.
For the sake of safety, it is recommended to employ AdaptGuard as a model preprocessing step to extract effective information within the source model and discard its risky parameters.
Then take the student network as initialization for their own model adaptation algorithm.
With no restrictions on source training or target adaptation and no requirements about the network structure, AdaptGuard can be a general solution applied to many scenarios.

\setlength{\tabcolsep}{3.0pt}
    \begin{table*}[!t]
        \centering
        \caption{Accuracies (\%) of various defense methods against four attacks on \textbf{DomainNet126} \cite{peng2019moment} dataset for model adaptation (ResNet-50).}
        \vspace{3mm}
        \resizebox{1.0\textwidth}{!}{
            \begin{tabular}{ll|cccca|cccca|cccca|cccca}
            \toprule
            \multicolumn{2}{c|}{Attack} & \multicolumn{5}{c|}{UAP \cite{moosavi2017universal}} & \multicolumn{5}{c|}{GAP \cite{poursaeed2018generative}} & \multicolumn{5}{c|}{Blended \cite{chen2017targeted}} & \multicolumn{5}{c}{SIG \cite{barni2019new}} \\
            \midrule
            \multicolumn{2}{c|}{Task} & CP & PR & RS & SC & Avg & CP & PR & RS & SC & Avg & CP & PR & RS & SC & Avg & CP & PR & RS & SC & Avg \\ 
            \midrule
            \multicolumn{2}{c|}{Clean (Source Only)} & 46.7 & 74.1 & 48.6 & 56.8 & 56.5 & 46.7 & 74.1 & 48.6 & 56.8 & 56.5 & 44.7 & 73.3 & 48.1 & 54.6 & 55.2 & 44.6 & 73.2 & 48.1 & 55.3 & 55.3 \\
            \multicolumn{2}{c|}{Attack (Source Only)} & 42.3 & 49.1 & 6.8 & 37.4 & 33.9 & 14.2 & 42.4 & 11.6 & 34.5 & 25.7 & 0.4 & 0.4 & 0.0 & 0.6 & 0.4 & 0.9 & 1.3 & 0.0 & 0.2 & 0.6  \\
            \midrule
            \midrule
            \multirow{4}{3.5em}{Clean (SHOT)} & \multicolumn{1}{|l|}{SHOT \cite{liang2020we}} & \bestclean{61.0} & \bestclean{79.8} & \bestclean{56.3} & \bestclean{69.8} & \bestclean{66.8} & \bestclean{61.0} & \bestclean{79.8} & \bestclean{56.3} & \bestclean{69.8} & \bestclean{66.8} & \bestclean{60.2} & \bestclean{79.9} & \bestclean{55.7} & \bestclean{69.4} & \bestclean{66.3} & \bestclean{60.5} & \bestclean{79.8} & 55.1 & \bestclean{69.4} & \bestclean{66.2}  \\
            & \multicolumn{1}{|l|}{+ANP \cite{wu2021adversarial}} & 52.9 & 78.6 & 49.7 & 66.9 & 62.0 & 52.9 & 78.6 & 49.7 & 66.9 & 62.0 & 52.5 & 77.4 & 52.5 & 64.8 & 61.8 & 53.5 & 76.9 & 52.3 & 66.4 & 62.3  \\
            & \multicolumn{1}{|l|}{+TRADES \cite{zhang2019theoretically}} & 57.7 & 41.7 & 53.2 & 69.4 & 55.5 & 57.7 & 41.7 & 53.2 & 69.4 & 55.5 & 56.7 & 35.6 & 51.2 & 68.9 & 53.1 & 56.9 & 53.4 & 51.7 & 68.3 & 57.6  \\
            & \multicolumn{1}{|l|}{+PGD \cite{madry2017towards}} & 54.7 & 66.1 & 53.0 & 68.4 & 60.5 & 54.7 & 66.1 & 53.0 & 68.4 & 60.5 & 53.8 & 40.1 & 53.9 & 69.0 & 54.2 & 54.2 & 37.9 & 53.3 & 68.0 & 53.4  \\
            & \multicolumn{1}{|l|}{+AdaptGuard} & 54.7 & 77.7 & 55.8 & 67.6 & 64.0 & 54.7 & 77.7 & 55.8 & 67.6 & 64.0 & 53.1 & 77.5 & \bestclean{55.7} & 66.0 & 63.1 & 53.1 & 77.6 & \bestclean{56.4} & 66.8 & 63.5  \\
            \midrule
            \multirow{4}{3em}{Attack (SHOT)} & \multicolumn{1}{|l|}{SHOT \cite{liang2020we}} & \best{58.8} & 61.7 & 41.4 & 67.9 & 57.4 & 39.3 & 51.0 & 37.3 & 57.5 & 46.3 & 31.9 & 54.5 & 30.6 & 38.0 & 38.7 & 47.2 & 71.6 & 47.1 & 54.4 & 55.1  \\
            & \multicolumn{1}{|l|}{+ANP \cite{wu2021adversarial}} & 50.8 & 64.9 & 39.2 & 65.1 & 55.0 & 38.2 & 48.3 & 35.7 & 57.8 & 45.0 & 35.9 & 50.2 & 35.5 & 49.8 & 42.9 & 50.7 & 72.8 & 47.1 & 56.7 & 56.8   \\
            & \multicolumn{1}{|l|}{+TRADES \cite{zhang2019theoretically}} & 57.0 & 42.4 & 52.8 & \best{69.1} & 55.3 & \best{55.2} & 43.2 & 52.1 & \best{67.9} & 54.6 & 37.4 & 21.5 & 41.2 & 55.9 & 39.0 & \best{53.8} & 49.8 & 50.6 & 66.3 & 55.1  \\
            & \multicolumn{1}{|l|}{+PGD \cite{madry2017towards}} & 54.2 & 65.3 & 52.8 & 68.2 & 60.1 & 53.2 & 64.7 & 52.3 & 67.8 & 59.5 & \best{39.9} & 26.4 & 42.3 & \best{59.1} & 41.9 & 52.7 & 35.7 & 52.2 & \best{67.4} & 52.0  \\
            & \multicolumn{1}{|l|}{+AdaptGuard} & 54.3 & \best{76.6} & \best{55.2} & 67.4 & \best{63.3} & 52.0 & \best{70.7} & \best{54.4} & 64.9 & \best{60.5} & 39.5 & \best{61.1} & \best{45.4} & 54.9 & \best{50.2} & 51.5 & \best{75.5} & \best{54.7} & 65.0 & \best{61.6}  \\
            \midrule
            \midrule
            \multirow{4}{3em}{Clean (NRC)} & \multicolumn{1}{|l|}{NRC \cite{yang2021exploiting}} & \bestclean{63.5} & \bestclean{82.0} & \bestclean{60.7} & \bestclean{72.1} & \bestclean{69.6} & \bestclean{63.5} & \bestclean{82.0} & \bestclean{60.7} & \bestclean{72.1} & \bestclean{69.6} & \bestclean{61.9} & \bestclean{81.5} & \bestclean{60.7} & \bestclean{71.6} & \bestclean{68.9} & \bestclean{62.5} & \bestclean{81.4} & \bestclean{60.8} & \bestclean{71.3} & \bestclean{69.4}  \\ 
            & \multicolumn{1}{|l|}{+ANP \cite{wu2021adversarial}} & 62.4 & 81.3 & 60.3 & 71.8 & 68.9 & 62.4 & 81.3 & 60.3 & 71.8 & 68.9 & 61.7 & 80.7 & 59.8 & 69.9 & 68.0 & 61.3 & 80.8 & 60.4 & 70.2 & 68.4  \\
            & \multicolumn{1}{|l|}{+TRADES \cite{zhang2019theoretically}} & 32.2 & 55.5 & 35.6 & 63.7 & 46.7 & 32.2 & 55.5 & 35.6 & 63.7 & 46.7 & 30.9 & 32.0 & 35.5 & 62.3 & 40.2 & 27.9 & 49.3 & 33.4 & 56.5 & 44.9  \\ 
            & \multicolumn{1}{|l|}{+PGD \cite{madry2017towards}} & 49.6 & 64.6 & 50.9 & 66.9 & 58.0 & 49.6 & 64.6 & 50.9 & 66.9 & 58.0 & 46.3 & 65.4 & 51.0 & 66.0 & 57.2 & 46.0 & 64.6 & 51.8 & 66.4 & 58.6  \\
            & \multicolumn{1}{|l|}{+AdaptGuard} & 56.8 & 78.1 & 59.8 & 67.7 & 65.6 & 56.8 & 78.1 & 59.8 & 67.7 & 65.6 & 55.2 & 78.3 & 60.5 & 66.2 & 65.0 & 55.5 & 78.0 & 60.6 & 66.3 & 65.3  \\ 
            \midrule
            \multirow{4}{3em}{Attack (NRC)} & \multicolumn{1}{|l|}{NRC \cite{yang2021exploiting}} & \best{61.7} & 59.0 & 38.4 & 69.3 & 57.1 & 40.4 & 47.2 & 33.9 & 58.5 & 45.0 & 36.6 & 44.5 & 34.0 & 48.6 & 40.9 & 51.4 & 65.2 & 32.4 & 36.4 & 45.8  \\ 
            & \multicolumn{1}{|l|}{+ANP \cite{wu2021adversarial}} & 60.5 & 62.7 & 41.1 & \best{69.7} & 58.5 & 41.7 & 53.2 & 36.0 & 59.5 & 47.6 & 37.5 & 45.1 & 36.6 & 48.8 & 42.0 & 56.7 & 73.5 & 44.6 & 60.2 & 54.3  \\ 
            & \multicolumn{1}{|l|}{+TRADES \cite{zhang2019theoretically}} & 32.1 & 47.9 & 34.4 & 63.2 & 44.4 & 22.0 & 37.2 & 31.2 & 60.9 & 37.8 & 10.6 & 7.6 & 21.1 & 43.4 & 20.7 & 7.1 & 16.5 & 24.7 & 39.2 & 25.3   \\ 
            & \multicolumn{1}{|l|}{+PGD \cite{madry2017towards}} & 49.4 & 64.3 & 50.7 & 66.7 & 57.8 & 47.3 & 64.0 & 50.0 & 66.1 & 56.9 & 31.8 & 52.0 & 41.9 & \best{59.5} & 46.3 & 39.4 & 62.4 & 50.5 & 65.0 & 53.8  \\
            & \multicolumn{1}{|l|}{+AdaptGuard} & 54.3 & \best{76.6} & \best{55.2} & 67.4 & \best{63.3} & \best{55.5} & \best{76.9} & \best{58.8} & \best{66.5} & \best{64.4} & \best{41.5} & \best{61.1} & \best{49.1} & 57.8 & \best{52.4} & \best{54.1} & \best{75.5} & \best{59.4} & \best{65.5} & \best{60.8}  \\
            \bottomrule
            \end{tabular}
        }
    \label{tab: result domainnet126}
    \end{table*}

\section{Experiment}

\subsection{Setup}

\textbf{Datasets.}
Our experiments are based on three popular image classification datasets in the model adaptation task.
\textbf{Office} \cite{saenko2010adapting} is a standard benchmark whose images are collected from the office environment.
It contains 31 categories across three domains (\ie, Amazon (A), DSLR (D), and Webcam (W)).
\textbf{OfficeHome} \cite{venkateswara2017deep} is a medium-size dataset consisting of 65 categories and four domains (\ie, Art (A), Clipart (C), Product (P), and Real World (R)). OfficeHome has a more balanced sample number across domains and is now the most used dataset.
\textbf{DomainNet126} \cite{saito2019semi}, a subset version of DomainNet \cite{peng2019moment}, is a large-size benchmark that consists of 126 categories and 4 domains (\ie, Clipart (C), Painting (P), Real (R), and Sketch (S)).
Different from the above datasets, it has a separate test set, and we only choose four tasks in a loop (\ie, CP, PR, RS, and SC) due to a large number of experiments.

\textbf{Implementation Details.}
We choose two popular model adaptation methods, SHOT \cite{liang2020we} and NRC \cite{yang2021exploiting}, as our base methods.
Also, we decide to use two universal adversarial perturbations (\ie, UAP \cite{moosavi2017universal} and GAP \cite{poursaeed2018generative}) and two backdoor attacks (\ie, Blended \cite{chen2017targeted} and SIG \cite{barni2019new}) as universal attack methods from the source domain.
As UAP and GAP generate perturbation based on a well-trained model, they share the same source models with the original model adaptation algorithms.
Blended and SIG embed the backdoor into the model using different patterns, so we train a set of source models for each one.
$L_{inf}$ norm of universal perturbation generated using UAP and GAP is required to be no greater than 10/255.
In Blended and SIG, the poisoning rate in the training dataset is set to 0.1 and we choose 0 as the target category in all experiments.
Note that as the dataset in model adaptation is relatively small, we select the source poison data from the whole categories in SIG.
Finally, we use the accuracies of the clean samples and attacked samples (\ie, Clean and Attack) as evaluation metrics to evaluate the security of the algorithms.

\textbf{Hyperparameters.}
We use the same hyperparameters of AdaptGuard in all experiments.
The default epoch number of knowledge distillation is 50. The weight of EMA progress in Eq.~(\ref{eq:kd_clean}) is set to 0.6.
And as shown in Eq.~(\ref{eq:method_kd_all}), both weights in loss functions are 0.5.
About PGD \cite{madry2017towards} algorithm, we search 7 iterations in the area whose $L_{inf}$ norm is less than 4/255 and calculate step size in the way recommended by PGD \cite{madry2017towards}.
For the hyperparameters in SHOT \cite{liang2020we} and NRC \cite{yang2021exploiting}, we follow their official settings.

\textbf{Baselines.}
Since there is no previous work about the universal attack on model adaptation tasks, we use several robust training strategies as comparison methods.
Considering backdoor defense, we use ANP \cite{wu2021adversarial} to cut the sensitive BatchNorm layers from the source model and then employ adaptation.
As ANP fails under unlabeled data, we use half of the clean source dataset in its optimization process, which is not allowed in our framework.
Besides, we add PGD \cite{madry2017towards} and TRADES \cite{zhang2019theoretically} terms to model adaptation methods as an additional loss function.
The trade-off hyperparameters of PGD and TRADES are sensitive in joint training, so we use 0.2 in Office and OfficeHome and 0.1 in DomainNet126 to ensure convergence.
In addition, SHOT can also be regarded as a fine-tune-based defense method.

\begin{figure*}[!htb]
		\centering
		\small
		\setlength\tabcolsep{1mm}
		\renewcommand\arraystretch{0.1}
		\begin{tabular}{cccc}
			\includegraphics[width=0.24\linewidth]{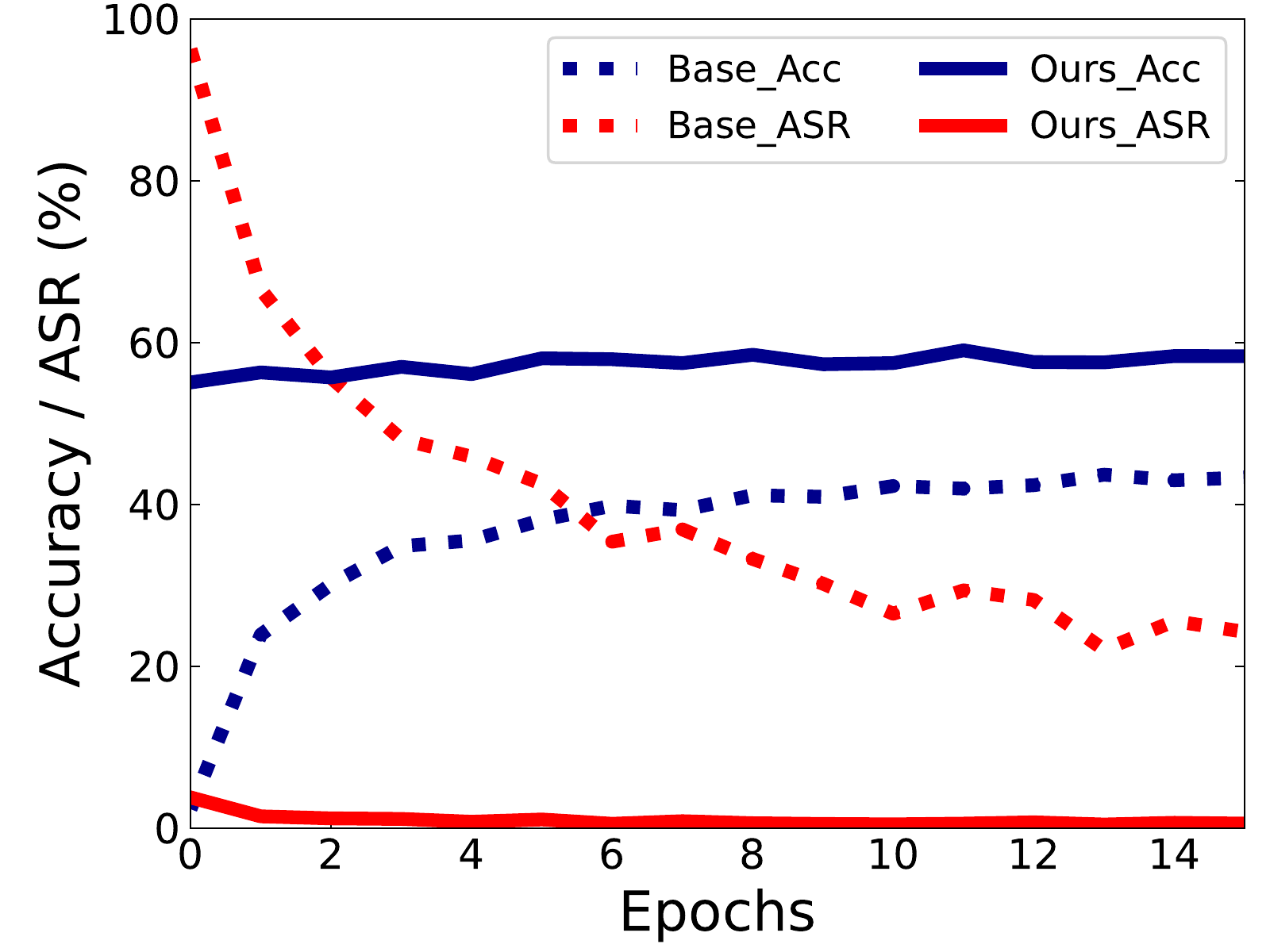} &
			\includegraphics[width=0.24\linewidth, clip]{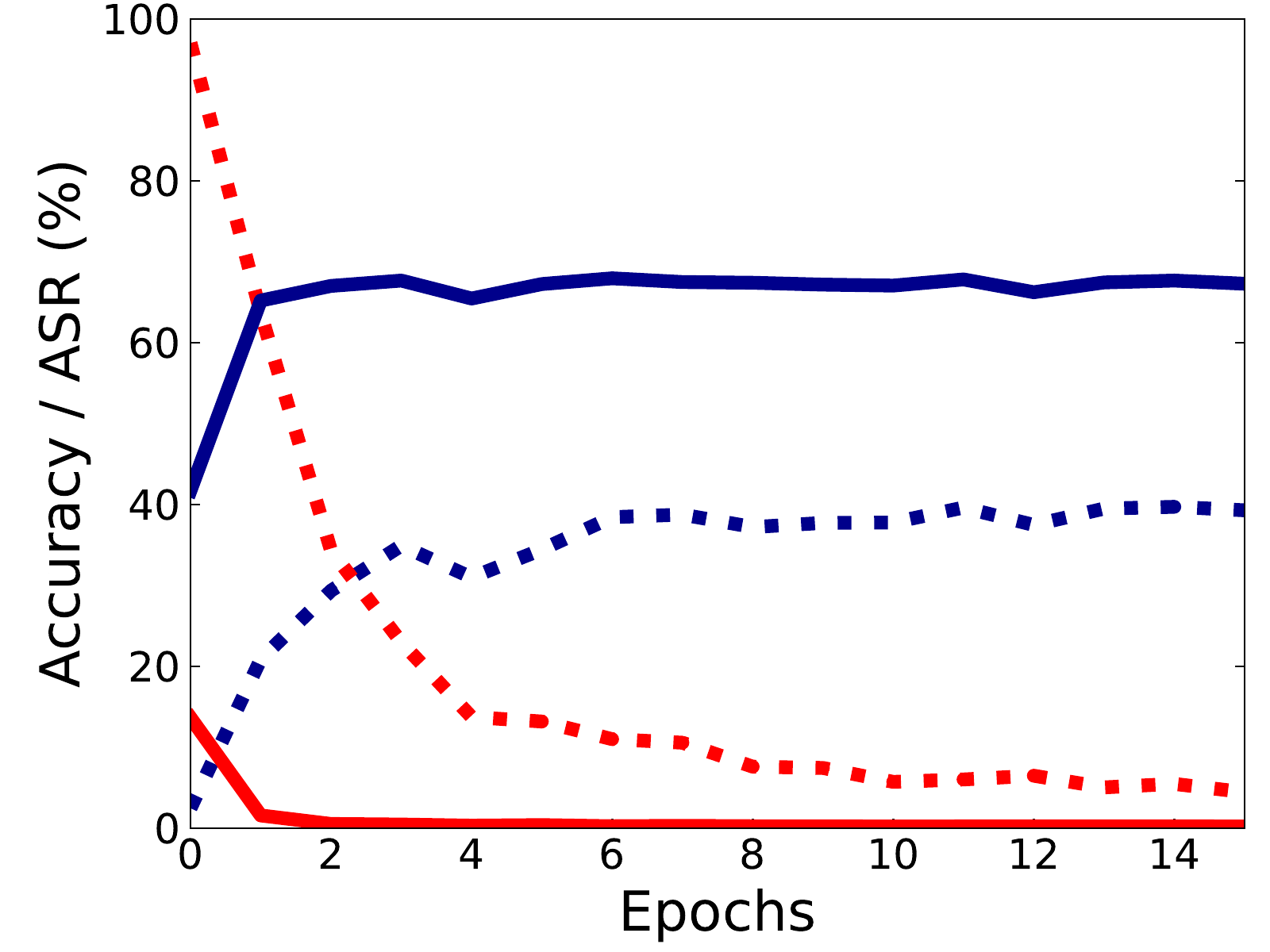} & 
			\includegraphics[width=0.24\linewidth, clip]{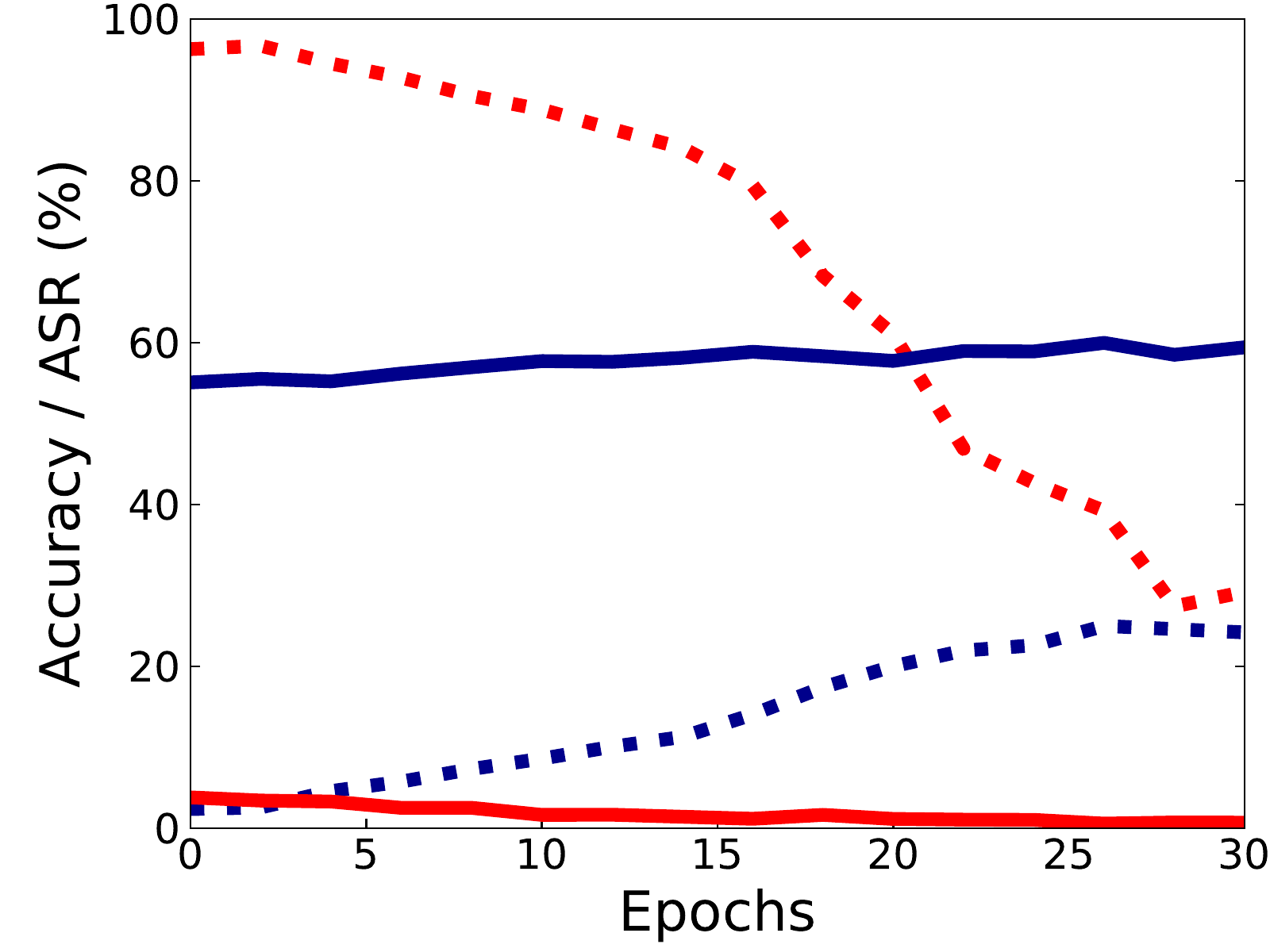} &
			\includegraphics[width=0.24\linewidth, clip]{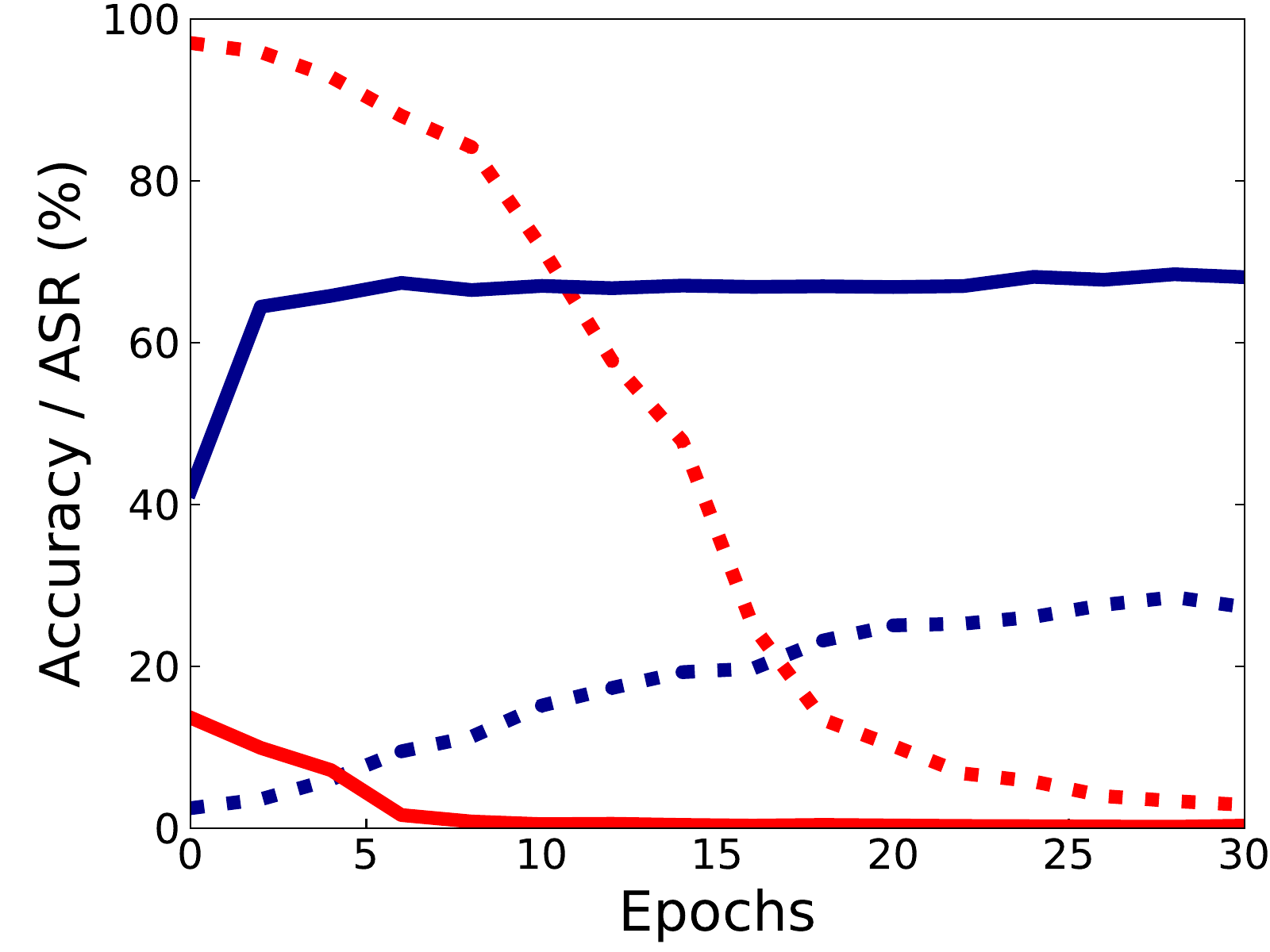} \\
			\\[0.5mm]
			(a) SHOT under Blended attack. & (b) SHOT under SIG attack.  & (c) NRC under Blended attack. & (d) NRC under SIG attack.
		\end{tabular}
        \vspace{1pt}
		\caption{Analysis of Performance trend of backdoor defense on \textbf{A$\to$P} from \textbf{OfficeHome} \cite{venkateswara2017deep}.}
		\label{fig:backdoor}
\end{figure*} 

\subsection{Results}
We evaluate AdaptGuard on three image classification datasets against four universal attacks and the results are shown in Table~\ref{tab: result office}, \ref{tab: result office-home}, \ref{tab: result domainnet126}.
Due to space limitations, for Office and OfficeHome, we only report the average value according to the source domain and leave the detailed results in the \textbf{supplementary material}.
And we also evaluate AdaptGuard under image-specific attacks from the source side to show the flexibility of our framework, whose results are provided in Table~\ref{tab: pgd-attack}.

\textbf{Analysis about universal adversarial perturbations.}
Universal adversarial perturbations generated from the source domain can transferred to the target domain.
As shown in Table~\ref{tab: result office-home}, the performance of the source-only model drops 41.5\% when attacked by UAP.
Model adaptation methods also have the ability to defend against perturbations.
When employing SHOT on the source-only model, this drop gap narrows to 27.0\%.
PGD and TRADES can provide an effective defense against adversarial perturbations, but they make a huge impact on the clean sample performance.
AdaptGuard achieves satisfactory defense against universal adversarial perturbations and maintains the transfer capability of target samples.

\textbf{Analysis about backdoor attacks.}
The backdoor embedded in the source model can achieve attacks on the target domain.
As shown in Table~\ref{tab: result domainnet126}, for source-only models, the accuracy on the poisoned images is less than 1\%, but performance on the clean test set is about 55\%.
Similar phenomena are also found in other benchmarks.
Model adaptation methods can defend against backdoor attacks to some extent.
It is found that SHOT achieves a better defense ability than NRC from the same victim source model.
We think SHOT employs fine-tuning with pseudo labels which is regarded as a direct backdoor defense while NRC only exploits the target domain structure.
PGD and TRADES can defend against backdoor attacks in some tasks.
They generate adversarial examples to provide more diverse samples for the victim model, promoting forgetting of training set knowledge, and weakening the effectiveness of backdoor attacks.
ANP acts as a good defense method against backdoor attacks, but may not be suitable for model preprocessing in transfer problems because pruning has side effects on the subsequent model adaptation methods.
From the experiment results, AdaptGuard defends against backdoor attacks better than compared methods.
Take OfficeHome as an example, as shown in Table~\ref{tab: result office-home}, AdaptGuard achieves 51.7\% and 62.4\% accuracy based on SHOT, and 50.3\% and 60.6\% accuracy based on NRC under Blend and SIG respectively.

\setlength{\tabcolsep}{4.0pt}
    \begin{table}[!t]
        \centering
        \caption{Accuracies (\%) of various defense methods against \textbf{image-specific attack} on three benchmarks for model adaptation.}
        \vspace{3mm}
        \resizebox{0.44\textwidth}{!}
        {
            \begin{tabular}{l|lll}
            \toprule
            Datasets & OfficeHome & Office & DomainNet126 \\
            \midrule
            SHOT \cite{liang2020we} & 13.7 & 11.1 & 26.1 \\
            +ANP \cite{wu2021adversarial} & 27.2 & 19.5 & 31.7 \\
            +TRADES \cite{zhang2019theoretically} & 61.6 & 57.1 & 52.5 \\
            +PGD \cite{madry2017towards} & 67.0 & 63.9 & 59.2 \\
            +AdaptGuard & \best{67.1} (+0.1) & \best{84.7} (+20.8) & \best{60.5} (+1.3) \\
            \midrule
            NRC \cite{yang2021exploiting} & 5.5 & 6.7 & 17.0 \\
            +ANP \cite{wu2021adversarial} & 16.1 & 9.9 & 24.3 \\
            +TRADES \cite{zhang2019theoretically} & 41.9 & 70.1 & 25.1 \\
            +PGD \cite{madry2017towards} & 57.5 & 83.8 & 55.6 \\
            +AdaptGuard & \best{65.6} (+8.1) & \best{85.5} (+1.7) & \best{63.3} (+7.7) \\
            \bottomrule
            \end{tabular}
        }   
    \label{tab: pgd-attack}
    \end{table}

\textbf{Analysis about accuracy and ASR during the backdoor defense.} Attack success rate (ASR) refers to the proportion of samples that can be classified into the target class after the backdoor attack.
It is a common metric for evaluating backdoor defenses.
We provide the curves of accuracy and ASR in the target during model adaptation shown in Fig.~\ref{fig:backdoor}.
For model adaptation methods, ASR drops rapidly at the beginning of training.
This phenomenon proves that more samples being misled to the target class are misclassified to other wrong classes.
These misclassified samples make ASR value better but contribute nothing to the adaptation task, so it is necessary to consider more accuracy metrics.
Our method, as shown in the curves, supports a robust initialization to downstream adaptation methods and achieves higher and stabler accuracy of attacked images.

\textbf{Defend against image-specific attack}. 
For a deeper discussion, we also validate AdaptGuard under the image-specific attack.
We use a PGD-7 \cite{madry2017towards} with $L_{inf}$ norm 4/255 from the source domain to generate an adversarial sample for each test instance, and report the attack accuracy of the target model in Table~\ref{tab: pgd-attack}.
From the results, the performance of model adaptation algorithms drops a lot under image-specific attacks from the source side, and AdaptGuard can also provide a strong defense in various benchmarks.
This implies that our framework has attractive flexibility.

\setlength{\tabcolsep}{2.0pt}
    \begin{table}[!t]
        \centering
        \caption{\textbf{Ablation studies} on three tasks (\ie, \textbf{A$\to$P} in OfficeHome, \textbf{A$\to$W} in Office, and \textbf{P$\to$R} in DomainNet126).}
        \vspace{3mm}
        \resizebox{0.5\textwidth}{!}
        {
            \begin{tabular}{c|l|ccc|ccc}
            \toprule
            \multicolumn{2}{c|}{Attack} & \multicolumn{3}{c|}{UAP \cite{moosavi2017universal}} & \multicolumn{3}{c}{Blended \cite{chen2017targeted}} \\
            \midrule
            \multicolumn{2}{c|}{Task} & A$\to$P & A$\to$W & P$\to$R & A$\to$P & A$\to$W & P$\to$R \\ 
            \midrule
            \multirow{4}{3em}{Clean} & SHOT \cite{liang2020we} & 78.0 & \bestclean{91.6} & 78.8 & \bestclean{77.9} & \bestclean{90.8} & 79.9 \\
            & +AdaptGuard (w/o adv) & \bestclean{78.2} & 88.7 & \bestclean{80.5} & 76.5 & 88.7 & \bestclean{80.4} \\
            & +AdaptGuard (w/o adjust) &  74.0 & 83.1 & 74.9 & 73.4 & 82.5 & 75.2 \\
            & +AdaptGuard & 77.0 & 83.9 & 77.7 & 76.8 & 84.0 & 77.5 \\
            \midrule
            \multirow{4}{3em}{Attack} & SHOT \cite{liang2020we} & 52.1 & 74.5 & 61.7 & 43.4 & 4.1 & 54.5 \\
            & +AdaptGuard (w/o adv) & 47.9 & 71.8 & 63.7 & 56.7 & 56.7 & \best{64.4} \\
            & +AdaptGuard (w/o adjust) & 73.7 & 76.7 & 74.1 & 57.1 & 59.4 & 59.9  \\
            & +AdaptGuard & \best{76.6} & \best{83.9} & \best{76.6} & \best{58.3} & \best{61.6} & 61.1 \\
            \bottomrule
            \end{tabular}
        }
    \label{tab: ablation}
    \end{table}

\subsection{Ablation Study}
To further study the contribution of terms in our proposed method, we investigate the effectiveness of introducing adversarial samples and the radius-adjusting strategy used in AdaptGuard.
When adversarial examples are not used, we adjust the weight of the loss function to 1.0 as shown in Eq.~(\ref{eq:kd_clean}), in order to ensure that the value of the loss function does not change much.
As shown in Table~\ref{tab: ablation}, AdaptGuard significantly improves the model's defense against backdoor attacks as a network preprocessing process.
The introduction of adversarial examples improves the adversarial robustness of the model but also affects the performance of clean samples.
The radius-adjusting strategy further enhances the defense ability against two kinds of attacks and mitigates the negative impact of adversarial examples on performance.

\section{Conclusion}
In this paper, we study the vulnerability of model adaptation methods toward universal attacks across domains. 
Existing model adaptation methods fail under the universal adversarial attack and the backdoor attack from the source model providers.
We propose a defense framework named AdaptGuard to defend against potential attacks from the source domain and maintain the target performance without requiring additional resources.
AdaptGuard employs knowledge distillation to avoid direct copying source parameters and utilizes adversarial examples with a gradually adjusted searching radius.
Experiments on common datasets validate that AdaptGuard can effectively defend against universal attacks in the model adaptation scenario.
And we plan to investigate various model adaptation tasks (e.g., segmentation) under more complex image-specific attacks (e.g., transfer-based attacks) in future work to extend our framework in broader application scenarios.

\section*{Acknowledgment}
This work was partially funded by National Natural Science Foundation of China under Grants (62276256 and U21B2045) and Beijing Nova Program under Grant Z211100002121108.
The authors would like to thank Zi Wang (AHU) and Jiyang Guan (CASIA) for their valuable discussions.

{\small
\bibliographystyle{ieee_fullname}
\bibliography{main}
}

\newpage

\appendix

\section{Experiment with more backdoor attacks}
Besides backdoor methods reported in the paper (\ie, Blended \cite{chen2017targeted} and SIG \cite{barni2019new}), we evaluate our method under the other two backdoor attacks (\ie, BadNets \cite{gu2017badnets} and WaNet \cite{nguyen2021wanet}) on OfficeHome dataset and results are shown in Table. \ref{tab: more backdoor}.
It is shown that AdaptGuard also works well under these backdoor attacks which indicates its versatility for attack methods.

\section{Detailed Experiment}
In this section, we provide detailed experimental results. 
Due to the space limitation of the paper, the results of some datasets are given in the form of source domain averages, and the detailed values are provided here.

\subsection{Universal Adversarial Perturbation}
We provide the detailed results about universal adversarial perturbation (\ie, UAP \cite{moosavi2017universal} and GAP \cite{poursaeed2018generative}) in Tables.~\ref{tab: result office-home UAP},~\ref{tab: result office UAP}.

\subsection{Backdoor Attack}
Different backdoor attack uses different source models, thus we report the results of each backdoor attack separately.
The results of Blended \cite{chen2017targeted} attack are listed in Tables.~\ref{tab: result office-home Blended},~\ref{tab: result office Blended}.
And we provide the results about SIG \cite{barni2019new} attack in Tables.~\ref{tab: result office-home SIG},~\ref{tab: result office SIG}.

\setlength{\tabcolsep}{3.0pt}
    \begin{table*}[!ht]
        \centering
        \caption{Accuracies and ASR (\%) of AdaptGuard against BadNets \cite{gu2017badnets} and WaNet \cite{nguyen2021wanet} on \textbf{OfficeHome} \cite{venkateswara2017deep} dataset for SHOT \cite{liang2020we} (ResNet-50). Best results are shown in \textbf{bold}.}
        \vspace{3mm}
        \resizebox{0.45\textwidth}{!}
        {
            \begin{tabular}{l|ccc|ccc}
            \toprule
            & \multicolumn{3}{c|}{BadNets \cite{gu2017badnets}} & \multicolumn{3}{c}{WaNet \cite{nguyen2021wanet}} \\
            & Clean & Attack & ASR & Clean & Attack & ASR \\
            \midrule
            SHOT \cite{liang2020we} & \textbf{70.7} & 15.5 & 80.2 & \textbf{70.9} & 35.0 & 48.3  \\
            +AdaptGuard & 67.4 & \textbf{66.1} & \textbf{2.3} & 67.8   & \textbf{65.8} & \textbf{2.6} \\
            \bottomrule
            \end{tabular}
        }
    \label{tab: more backdoor}
    \vspace{-20mm}
    \end{table*}

\setlength{\tabcolsep}{3.0pt}
    \begin{table*}[!t]
        \centering
        \caption{Accuracies (\%) of defense methods against universal adversarial perturbation attacks on \textbf{OfficeHome} \cite{venkateswara2017deep} dataset for model adaptation (ResNet-50). Best accuracy of clean samples (\bestclean{bold}), best accuracy of attacked samples (\best{bold red}).}
        \vspace{3mm}
        \resizebox{1.0\textwidth}{!}{
            \begin{tabular}{ll|cccccccccccca}
            \toprule
            \multicolumn{2}{c|}{Task} & A$\to$C & A$\to$P & A$\to$R & C$\to$A & C$\to$P & C$\to$R & P$\to$A & P$\to$C & P$\to$R & R$\to$A & R$\to$C & R$\to$P & Avg \\ 
            \midrule
            \multicolumn{2}{c|}{Clean (Source Only)} & 43.9 & 65.3 & 72.9 & 52.9 & 60.7 & 64.4 & 51.1 & 40.8 & 73.1 & 64.9 & 45.2 & 77.1 & 59.4 \\
            \multicolumn{2}{c|}{\textbf{UAP} \cite{moosavi2017universal} Attack (Source Only)} & 20.7 & 32.4 & 36.1 & 11.4 & 13.2 & 16.8 & 12.0 & 8.9 & 16.4 & 16.2 & 11.9 & 18.6 & 17.9 \\
            \multicolumn{2}{c|}{\textbf{GAP} \cite{poursaeed2018generative} Attack (Source Only)} & 28.6 & 39.2 & 43.2 & 14.0 & 16.1 & 18.3 & 13.4 & 13.8 & 21.5 & 18.5 & 17.3 & 30.2 & 22.8 \\
            \midrule
            \midrule
            \multirow{5}{3.6em}{Clean (SHOT)} & \multicolumn{1}{|l|}{SHOT \cite{liang2020we}} & 56.5 & \bestclean{78.0} & \bestclean{81.3} & \bestclean{67.6} & \bestclean{77.4} & \bestclean{78.8} & \bestclean{68.2} & \bestclean{54.9} & \bestclean{82.6} & \bestclean{74.0} & 58.6 & \bestclean{84.4} & \bestclean{71.9} \\
            & \multicolumn{1}{|l|}{+ANP \cite{wu2021adversarial}} & 54.5 & 76.4 & 79.2 & 66.3 & 76.5 & 75.7 & 64.4 & 54.3 & 81.9 & 72.9 & \bestclean{58.7} & 83.2 & 70.3 \\
            & \multicolumn{1}{|l|}{+TRADES \cite{zhang2019theoretically}} & \bestclean{57.1} & 76.5 & 78.1 & 67.1 & 75.2 & 76.8 & 67.2 & 53.8 & 81.6 & 73.7 & 58.4 & 81.6 & 70.6 \\
            & \multicolumn{1}{|l|}{+PGD \cite{madry2017towards}} & 54.6 & 74.6 & 77.6 & 66.8 & 73.8 & 75.3 & 65.9 & 52.1 & 80.2 & 73.3 & 56.2 & 81.0 & 69.3  \\
            & \multicolumn{1}{|l|}{+AdaptGuard} & 54.1 & 77.0 & 79.4 & 63.0 & 75.0 & 74.7 & 62.9 & 51.5 & 80.5 & 67.4 & 56.0 & 81.4 & 68.6 \\
            \midrule
            \multirow{5}{3.6em}{\textbf{UAP} \cite{moosavi2017universal} Attack (SHOT)} & \multicolumn{1}{|l|}{SHOT \cite{liang2020we}} & 46.4 & 52.1 & 53.3 & 52.6 & 48.0 & 58.0 & 45.4 & 30.3 & 45.4 & 34.9 & 37.3 & 34.7 & 44.9  \\
            & \multicolumn{1}{|l|}{+ANP \cite{wu2021adversarial}} & 49.2 & 59.0 & 59.3 & 54.4 & 56.0 & 57.0 & 40.8 & 35.4 & 49.6 & 36.1 & 40.5 & 38.3 & 48.0 \\
            & \multicolumn{1}{|l|}{+TRADES \cite{zhang2019theoretically}} & \best{56.9} & 75.9 & 76.5 & 62.8 & \best{74.8} & \best{74.9} & 61.4 & \best{54.1} & \best{79.9} & 61.2 & \best{58.2} & 80.2 & 68.1 \\
            & \multicolumn{1}{|l|}{+PGD \cite{madry2017towards}} & 54.6 & 74.7 & 77.1 & \best{65.4} & 73.7 & 74.6 & \best{63.5} & 52.1 & 79.3 & \best{67.7} & 56.2 & \best{80.7} & \best{68.3} \\
            & \multicolumn{1}{|l|}{+AdaptGuard} & 54.1 & \best{76.6} & \best{79.1} & 61.7 & 74.4 & 74.3 & 60.9 & 51.3 & 79.8 & 65.5 & 55.4 & 80.6 & 67.8 \\
            \midrule
            \multirow{5}{3.6em}{\textbf{GAP} \cite{poursaeed2018generative} Attack (SHOT)} & \multicolumn{1}{|l|}{SHOT \cite{liang2020we}} & 42.9 & 43.5 & 56.4 & 41.9 & 41.4 & 49.2 & 41.3 & 29.4 & 43.5 & 36.1 & 33.7 & 26.2  & 40.5 \\
            & \multicolumn{1}{|l|}{+ANP \cite{wu2021adversarial}} & 46.9 & 54.7 & 57.0 & 44.8 & 49.9 & 49.1 & 35.9 & 32.1 & 45.2 & 40.1 & 38.1 & 35.8  & 44.1 \\
            & \multicolumn{1}{|l|}{+TRADES \cite{zhang2019theoretically}} & \best{56.9} & \best{75.4} & 74.2 & 56.2 & \best{73.8} & 72.5 & 55.7 & \best{53.9} & 75.6 & 54.5 & 58.1 & 79.4  & 65.5 \\
            & \multicolumn{1}{|l|}{+PGD \cite{madry2017towards}} &  54.6 & 74.6 & 76.5 & \best{63.1} & 73.4 & \best{74.0} & \best{58.1} & 52.2 & \best{78.3} & 61.1 & \best{56.2} & \best{80.5}  & \best{66.9} \\
            & \multicolumn{1}{|l|}{+AdaptGuard} & 53.8 & 74.8 & \best{77.3} & 58.7 & 72.9 & 72.6 & 56.6 & 50.4 & 77.3 & \best{61.3} & 54.5 & 77.2 & 65.6 \\

            \midrule
            \midrule
            
            \multirow{5}{3.6em}{Clean (NRC)} & \multicolumn{1}{|l|}{NRC \cite{yang2021exploiting}} & \bestclean{57.6} & \bestclean{78.2} & \bestclean{81.4} & \bestclean{67.2} & \bestclean{78.5} & \bestclean{79.4} & \bestclean{64.8} & \bestclean{57.2} & \bestclean{81.4} & \bestclean{72.3} & 58.2 & \bestclean{85.1} & \bestclean{71.8} \\
            & \multicolumn{1}{|l|}{+ANP \cite{wu2021adversarial}} & 56.6 & 77.0 & 79.9 & 65.6 & 76.3 & 78.6 & 64.2 & 56.3 & 80.5 & 71.4 & \bestclean{58.5} & 84.1 & 70.7 \\
            & \multicolumn{1}{|l|}{+TRADES \cite{zhang2019theoretically}} & 50.5 & 72.0 & 72.2 & 59.5 & 71.0 & 66.8 & 55.7 & 47.7 & 72.8 & 67.4 & 52.6 & 78.3 & 63.9 \\
            & \multicolumn{1}{|l|}{+PGD \cite{madry2017towards}} &  50.0 & 72.5 & 72.7 & 64.7 & 70.6 & 69.8 & 61.4 & 49.1 & 74.3 & 69.2 & 53.4 & 77.0 & 65.4 \\
            & \multicolumn{1}{|l|}{+AdaptGuard} & 56.0 & 77.6 & 79.2 & 61.7 & 75.4 & 75.4 & 61.2 & 54.3 & 77.9 & 65.9 & 56.6 & 81.8 & 68.6 \\
            \midrule

            \multirow{5}{3.6em}{\textbf{UAP} \cite{moosavi2017universal} Attack (NRC)} & \multicolumn{1}{|l|}{NRC \cite{yang2021exploiting}} & 43.7 & 43.4 & 49.7 & 45.6 & 44.1 & 52.7 & 34.7 & 25.1 & 39.7 & 30.0 & 29.3 & 28.2 & 38.8  \\
            & \multicolumn{1}{|l|}{+ANP \cite{wu2021adversarial}} & 47.6 & 51.1 & 55.7 & 46.9 & 52.4 & 56.8 & 34.5 & 31.9 & 45.1 & 30.4 & 32.2 & 30.9 & 42.9 \\
            & \multicolumn{1}{|l|}{+TRADES \cite{zhang2019theoretically}} & 48.8 & 69.7 & 66.9 & 47.6 & 69.0 & 59.5 & 41.2 & 46.0 & 64.4 & 44.3 & 50.0 & 73.5 & 56.7 \\
            & \multicolumn{1}{|l|}{+PGD \cite{madry2017towards}} &  49.6 & 71.8 & 71.0 & 58.0 & 69.8 & 68.3 & 53.1 & 48.5 & 72.4 & 58.2 & 53.1 & 75.9 & 62.5 \\
            & \multicolumn{1}{|l|}{+AdaptGuard} & \best{55.5} & \best{76.9} & \best{78.8} & \best{58.8} & \best{74.6} & \best{72.3} & \best{58.6} & \best{53.8} & \best{77.4} & \best{63.6} & \best{56.0} & \best{80.9} & \best{67.3} \\
            \midrule

            \multirow{5}{3.6em}{\textbf{GAP} \cite{poursaeed2018generative} Attack (NRC)} & \multicolumn{1}{|l|}{NRC \cite{yang2021exploiting}} & 38.4 & 41.9 & 54.1 & 37.0 & 38.6 & 44.9 & 33.9 & 26.8 & 40.7 & 36.2 & 28.8 & 29.5 & 37.6 \\
            & \multicolumn{1}{|l|}{+ANP \cite{wu2021adversarial}} & 44.8 & 49.0 & 55.0 & 39.0 & 45.1 & 48.0 & 33.5 & 29.1 & 42.9 & 39.0 & 31.7 & 36.4 & 41.1 \\
            & \multicolumn{1}{|l|}{+TRADES \cite{zhang2019theoretically}} & 48.0 & 67.7 & 63.1 & 39.9 & 65.4 & 53.1 & 34.4 & 42.3 & 57.8 & 41.8 & 47.9 & 70.1 & 52.6 \\
            & \multicolumn{1}{|l|}{+PGD \cite{madry2017towards}} & 49.6 & 71.1 & 69.5 & 50.0 & 69.1 & 67.2 & 45.0 & 48.1 & 69.7 & 50.1 & 52.9 & 75.2 & 59.8  \\
            & \multicolumn{1}{|l|}{+AdaptGuard} & \best{54.8} & \best{75.8} & \best{77.9} & \best{54.8} & \best{73.5} & \best{71.8} & \best{54.9} & \best{52.9} & \best{76.7} & \best{59.2} & \best{55.1} & \best{78.8} & \best{65.5} \\
            
            \bottomrule
            \end{tabular}
        }
    \label{tab: result office-home UAP} 
    \end{table*}

\setlength{\tabcolsep}{4.0pt}
    \begin{table*}[!t]
        \centering
        \caption{Accuracies (\%) of defense methods against universal adversarial perturbation attacks on \textbf{Office} \cite{saenko2010adapting} dataset for model adaptation (ResNet-50).}
        \vspace{3mm}
        \resizebox{0.7\textwidth}{!}{
            \begin{tabular}{ll|cccccca}
            \toprule
            \multicolumn{2}{c|}{Task} & A$\to$D & A$\to$W & D$\to$A & D$\to$W & W$\to$A & W$\to$D & Avg \\ 
            \midrule
            \multicolumn{2}{c|}{Clean (Source Only)} & 82.1 & 75.5 & 61.6 & 95.2 & 62.5 & 97.6 & 79.1 \\
            \multicolumn{2}{c|}{\textbf{UAP} \cite{moosavi2017universal} Attack (Source Only)} & 38.6 & 36.6 & 43.4 & 50.7 & 36.4 & 48.6 & 42.4 \\
            \multicolumn{2}{c|}{\textbf{GAP} \cite{poursaeed2018generative} Attack (Source Only)} & 14.7 & 11.6 & 40.7 & 52.1 & 35.8 & 50.6 & 34.2 \\
            \midrule
            \midrule
            \multirow{5}{3.6em}{Clean (SHOT)} & \multicolumn{1}{|l|}{SHOT \cite{liang2020we}} & 93.6 & \bestclean{91.6} & \bestclean{74.6} & 97.2 & \bestclean{76.1} & \bestclean{99.8} & \bestclean{88.8} \\
            & \multicolumn{1}{|l|}{+ANP \cite{wu2021adversarial}} & 92.2 & 89.3 & 73.6 & \bestclean{98.1} & 75.2 & 99.0 & 87.9 \\
            & \multicolumn{1}{|l|}{+TRADES \cite{zhang2019theoretically}} & 92.0 & 90.1 & 74.3 & 97.2 & 75.1 & 98.4 & 87.8 \\
            & \multicolumn{1}{|l|}{+PGD \cite{madry2017towards}} & \bestclean{94.6} & 91.1 & 73.2 & 97.6 & 72.8 & \bestclean{99.8} & 88.2 \\
            & \multicolumn{1}{|l|}{+AdaptGuard} & 92.6 & 83.9 & 73.0 & 94.5 & 73.4 & 97.4 & 85.8 \\
            \midrule
            
            \multirow{5}{3.6em}{\textbf{UAP} \cite{moosavi2017universal} Attack (SHOT)} & \multicolumn{1}{|l|}{SHOT \cite{liang2020we}} & 66.7 & 74.5 & 67.4 & 76.6 & 67.6 & 62.3 & 69.2  \\
            & \multicolumn{1}{|l|}{+ANP \cite{wu2021adversarial}} & 66.3 & 76.9 & 66.2 & 76.7 & 68.2 & 66.7 & 70.2 \\
            & \multicolumn{1}{|l|}{+TRADES \cite{zhang2019theoretically}} & 82.9 & \best{87.9} & \best{73.8} & \best{96.4} & \best{74.7} & 87.8 & 83.9 \\
            & \multicolumn{1}{|l|}{+PGD \cite{madry2017towards}} & 87.4 & 87.8 & 73.1 & \best{96.4} & 72.9 & 91.0 & 84.7 \\
            & \multicolumn{1}{|l|}{+AdaptGuard} & \best{89.0} & 83.9 & 72.8 & 94.7 & 73.3 & \best{97.0} & \best{85.1} \\
            \midrule
            \multirow{5}{3.6em}{\textbf{GAP} \cite{poursaeed2018generative} Attack (SHOT)} & \multicolumn{1}{|l|}{SHOT \cite{liang2020we}} & 35.7 & 45.2 & 59.0 & 77.0 & 63.4 & 65.1 & 57.6  \\
            & \multicolumn{1}{|l|}{+ANP \cite{wu2021adversarial}} & 38.4 & 49.9 & 56.0 & 73.5 & 64.5 & 64.7 & 57.8 \\
            & \multicolumn{1}{|l|}{+TRADES \cite{zhang2019theoretically}} & 58.4 & \best{75.5} & \best{73.1} & 94.1 & \best{74.1} & 78.7 & 75.6 \\
            & \multicolumn{1}{|l|}{+PGD \cite{madry2017towards}} & \best{58.8} & 72.1 & 73.0 & 93.0 & 72.6 & 85.5 & 75.8  \\
            & \multicolumn{1}{|l|}{+AdaptGuard} & 49.2 & 74.3 & 72.7 & \best{94.5} & 72.9 & \best{92.8} & \best{76.1} \\

            \midrule
            \midrule

            \multirow{5}{3.6em}{Clean (NRC)} & \multicolumn{1}{|l|}{NRC \cite{yang2021exploiting}} & 92.8 & 89.1 & \bestclean{75.2} & 97.2 & \bestclean{74.4} & \bestclean{99.8} & 88.1 \\
            & \multicolumn{1}{|l|}{+ANP \cite{wu2021adversarial}} & 92.6 & 90.4 & 74.7 & 97.6 & 74.0 & \bestclean{99.8} & \bestclean{88.2} \\
            & \multicolumn{1}{|l|}{+TRADES \cite{zhang2019theoretically}} & 88.8 & 89.8 & 71.9 & \bestclean{97.9} & 69.9 & 98.2 & 86.1 \\
            & \multicolumn{1}{|l|}{+PGD \cite{madry2017towards}} & 93.2 & \bestclean{92.1} & 68.3 & 97.4 & 68.7 & 99.6 & 86.5 \\
            & \multicolumn{1}{|l|}{+AdaptGuard} & \bestclean{94.8} & 87.3 & 73.6 & 97.0 & 72.5 & 98.0 & 87.2 \\
            \midrule

            \multirow{5}{3.6em}{\textbf{UAP} \cite{moosavi2017universal} Attack (NRC)} & \multicolumn{1}{|l|}{NRC \cite{yang2021exploiting}} &  58.4 & 65.5 & 64.7 & 62.9 & 63.3 & 56.4 & 61.9 \\
            & \multicolumn{1}{|l|}{+ANP \cite{wu2021adversarial}} & 60.4 & 68.4 & 64.9 & 63.5 & 59.7 & 61.2 & 63.0  \\
            & \multicolumn{1}{|l|}{+TRADES \cite{zhang2019theoretically}} & 85.1 & 88.7 & 71.4 & 96.9 & 69.4 & 90.8 & 83.7 \\
            & \multicolumn{1}{|l|}{+PGD \cite{madry2017towards}} & \best{89.4} & \best{90.8} & 68.2 & \best{97.2} & 68.6 & \best{97.6} & \best{85.3}  \\
            & \multicolumn{1}{|l|}{+AdaptGuard} & 89.0 & 83.9 & \best{72.8} & 94.7 & \best{73.3} & 97.0 & 85.1 \\
            \midrule
            
            \multirow{5}{3.6em}{\textbf{GAP} \cite{poursaeed2018generative} Attack (NRC)} & \multicolumn{1}{|l|}{NRC \cite{yang2021exploiting}} & 31.3 & 39.9 & 61.2 & 65.7 & 60.5 & 60.4 & 53.2 \\
            & \multicolumn{1}{|l|}{+ANP \cite{wu2021adversarial}} & 33.9 & 44.7 & 58.2 & 60.0 & 58.5 & 60.6 & 52.6 \\
            & \multicolumn{1}{|l|}{+TRADES \cite{zhang2019theoretically}} & 66.5 & \best{83.1} & 70.8 & 95.0 & 68.6 & 81.7 & 77.6 \\
            & \multicolumn{1}{|l|}{+PGD \cite{madry2017towards}} &  \best{67.7} & 81.3 & 68.0 & 95.0 & 68.3 & \best{91.4} & \best{78.6} \\
            & \multicolumn{1}{|l|}{+AdaptGuard} & 49.8 & 77.4 & \best{72.1} & \best{96.1} & \best{71.4} & 89.6 & 76.1 \\
            
            \bottomrule
            \end{tabular}
        }
    \label{tab: result office UAP} 
    \end{table*}

\setlength{\tabcolsep}{3.0pt}
    \begin{table*}[!t]
        \centering
        \caption{Accuracies (\%) of defense methods against \textbf{Blended} \cite{chen2017targeted} backdoor attack on \textbf{OfficeHome} \cite{venkateswara2017deep} for model adaptation (ResNet-50).}
        \vspace{3mm}
        \resizebox{1.0\textwidth}{!}{
            \begin{tabular}{ll|cccccccccccca}
            \toprule
            \multicolumn{2}{c|}{Task} & A$\to$C & A$\to$P & A$\to$R & C$\to$A & C$\to$P & C$\to$R & P$\to$A & P$\to$C & P$\to$R & R$\to$A & R$\to$C & R$\to$P & Avg \\              
            \midrule
            \multicolumn{2}{c|}{Clean (Source Only)} & 44.3 & 64.5 & 72.9 & 47.4 & 59.2 & 61.3 & 50.4 & 39.8 & 71.7 & 62.4 & 45.0 & 75.4 & 57.9 \\
            \multicolumn{2}{c|}{\textbf{Blended} \cite{chen2017targeted} Attack (Source Only)} & 1.5 & 2.7 & 4.2 & 1.0 & 0.5 & 1.7 & 0.6 & 0.5 & 1.6 & 1.2 & 0.7 & 1.3 & 1.4 \\
            \midrule
            \midrule
        
            \multirow{5}{5.0em}{Clean (SHOT)} & \multicolumn{1}{|l|}{SHOT \cite{liang2020we}} & 56.3 & \bestclean{77.9} & \bestclean{80.4} & 66.1 & \bestclean{75.7} & \bestclean{77.6} & 67.1 & \bestclean{53.8} & \bestclean{81.9} & \bestclean{73.9} & \bestclean{59.3} & \bestclean{83.7} & \bestclean{71.1} \\
            & \multicolumn{1}{|l|}{+ANP \cite{wu2021adversarial}} & 53.2 & 77.1 & 78.9 & 62.1 & 66.6 & 71.0 & 63.2 & 52.9 & 80.9 & 2.2 & 2.2 & 1.7 & 51.0 \\
            & \multicolumn{1}{|l|}{+TRADES \cite{zhang2019theoretically}} & \bestclean{56.6} & 75.4 & 79.0 & 66.4 & 73.7 & 76.9 & 66.3 & \bestclean{53.8} & 81.3 & 73.6 & 57.7 & 81.7 & 70.2 \\
            & \multicolumn{1}{|l|}{+PGD \cite{madry2017towards}} & 54.2 & 74.9 & 77.2 & \bestclean{66.7} & 71.5 & 73.9 & \bestclean{67.2} & 51.6 & 79.6 & 73.4 & 56.5 & 80.9 & 68.9 \\
            & \multicolumn{1}{|l|}{+AdaptGuard} & 54.6 & 76.8 & 78.9 & 62.6 & 71.7 & 73.9 & 63.4 & 50.8 & 79.0 & 68.5 & 57.0 & 80.3 & 68.1 \\
            \midrule
       
            \multirow{5}{5.0em}{\textbf{Blended} \cite{chen2017targeted} Attack (SHOT)} & \multicolumn{1}{|l|}{SHOT \cite{liang2020we}} & 20.1 & 43.4 & 48.1 & 33.2 & 17.6 & 43.9 & 23.4 & 13.7 & 46.1 & 28.6 & 17.8 & 30.9 & 30.6 \\
            & \multicolumn{1}{|l|}{+ANP \cite{wu2021adversarial}} & 20.6 & 44.8 & 48.4 & 37.6 & 27.0 & 50.0 & 24.4 & 19.2 & 45.9 & 0.6 & 1.8 & 0.8 & 26.7 \\
            & \multicolumn{1}{|l|}{+TRADES \cite{zhang2019theoretically}} & 36.3 & 56.4 & 52.2 & 37.1 & 51.2 & 51.0 & 30.7 & 31.5 & 51.3 & 33.0 & 34.9 & 53.9 & 43.3  \\
            & \multicolumn{1}{|l|}{+PGD \cite{madry2017towards}} & \best{40.3} & 58.2 & 59.3 & 45.5 & \best{55.0} & 55.9 & 38.6 & \best{37.7} & 61.9 & 40.4 & \best{42.5} & \best{61.6} & 49.7 \\
            & \multicolumn{1}{|l|}{+AdaptGuard} & 38.1 & \best{58.3} & \best{63.9} & \best{49.1} & 53.4 & \best{59.5} & \best{45.1} & 34.4 & \best{63.8} & \best{53.3} & 41.3 & 60.7 & \best{51.7} \\
            \midrule
            \midrule
        
            \multirow{5}{5.0em}{Clean (NRC)} & \multicolumn{1}{|l|}{NRC \cite{yang2021exploiting}} & \bestclean{56.9} & \bestclean{78.2} & \bestclean{79.8} & \bestclean{66.2} & \bestclean{78.2} & \bestclean{78.4} & \bestclean{64.4} & \bestclean{57.1} & \bestclean{81.4} & \bestclean{71.4} & \bestclean{58.1} & \bestclean{83.2} & \bestclean{71.1} \\
            & \multicolumn{1}{|l|}{+ANP \cite{wu2021adversarial}} & 55.3 & 76.9 & 79.1 & 62.9 & 75.0 & 74.7 & 63.5 & 55.8 & 80.2 & 5.4 & 4.9 & 7.2 & 53.4 \\
            & \multicolumn{1}{|l|}{+TRADES \cite{zhang2019theoretically}} & 50.4 & 71.6 & 71.8 & 57.4 & 68.1 & 66.5 & 53.7 & 47.4 & 72.5 & 67.0 & 50.8 & 77.4 & 62.9 \\
            & \multicolumn{1}{|l|}{+PGD \cite{madry2017towards}} & 50.6 & 70.7 & 71.5 & 64.0 & 69.3 & 69.0 & 62.4 & 47.8 & 74.3 & 69.2 & 50.8 & 76.1 & 64.6  \\
            & \multicolumn{1}{|l|}{+AdaptGuard} & 56.2 & 76.8 & 79.0 & 60.3 & 72.6 & 72.1 & 62.4 & 51.7 & 76.8 & 65.7 & 56.6 & 80.3 & 67.5 \\
            \midrule
            \multirow{5}{5.0em}{\textbf{Blended} \cite{chen2017targeted} Attack (NRC)} & \multicolumn{1}{|l|}{NRC \cite{yang2021exploiting}} & 18.1 & 24.2 & 33.2 & 32.6 & 24.2 & 41.2 & 25.4 & 10.9 & 30.3 & 25.7 & 15.4 & 19.4 & 25.0  \\
            & \multicolumn{1}{|l|}{+ANP \cite{wu2021adversarial}} & 19.4 & 31.9 & 35.1 & 34.3 & 32.8 & 42.0 & 28.9 & 13.8 & 31.2 & 2.5 & 2.3 & 3.3 & 23.1 \\
            & \multicolumn{1}{|l|}{+TRADES \cite{zhang2019theoretically}} & 25.6 & 44.2 & 33.4 & 21.9 & 40.1 & 29.3 & 17.0 & 19.4 & 29.8 & 22.1 & 23.0 & 38.4 & 28.7 \\
            & \multicolumn{1}{|l|}{+PGD \cite{madry2017towards}} & 37.7 & 56.5 & 48.1 & 34.0 & 50.2 & 45.4 & 25.6 & 31.6 & 44.4 & 25.3 & 34.4 & 53.4 & 40.5 \\
            & \multicolumn{1}{|l|}{+AdaptGuard} & \best{38.8} & \best{59.4} & \best{60.8} & \best{44.5} & \best{53.7} & \best{56.4} & \best{44.5} & \best{33.0} & \best{59.3} & \best{51.7} & \best{39.5} & \best{61.4} & \best{50.3} \\
            
            \bottomrule
            \end{tabular}
        }
    \label{tab: result office-home Blended} 
    \end{table*}

\setlength{\tabcolsep}{4.0pt}
    \begin{table*}[!t]
        \centering
        \caption{Accuracies (\%) of defense methods against \textbf{Blended} \cite{chen2017targeted} backdoor attack on \textbf{Office} \cite{saenko2010adapting} dataset for model adaptation (ResNet-50).}
        \vspace{3mm}
        \resizebox{0.7\textwidth}{!}{
            \begin{tabular}{ll|cccccca}
            \toprule
            \multicolumn{2}{c|}{Task} & A$\to$D & A$\to$W & D$\to$A & D$\to$W & W$\to$A & W$\to$D & Avg \\ 
            \midrule
                   
            \multicolumn{2}{c|}{Clean (Source Only)} & 78.9 & 74.3 & 60.0 & 94.0 & 62.3 & 97.6 & 77.9 \\
            \multicolumn{2}{c|}{\textbf{Blended} \cite{chen2017targeted} Attack (Source Only)} & 0.4 & 0.3 & 1.2 & 0.0 & 0.5 & 2.1 & 0.7 \\

            \midrule
            \midrule
            \multirow{5}{5.0em}{Clean (SHOT)} & \multicolumn{1}{|l|}{SHOT \cite{liang2020we}} & \bestclean{94.6} & 90.8 & \bestclean{74.7} & 96.2 & \bestclean{74.8} & \bestclean{99.0} & \bestclean{88.4} \\
            & \multicolumn{1}{|l|}{+ANP \cite{wu2021adversarial}} & 91.8 & 88.3 & 73.7 & \bestclean{96.7} & 68.8 & \bestclean{99.0} & 86.4 \\
            & \multicolumn{1}{|l|}{+TRADES \cite{zhang2019theoretically}} & 92.8 & 90.1 & 74.0 & 95.7 & 73.7 & 97.6 & 87.3 \\            
            & \multicolumn{1}{|l|}{+PGD \cite{madry2017towards}} & 93.8 & \bestclean{91.2} & 72.5 & \bestclean{96.7} & 71.9 & \bestclean{99.0} & 87.5 \\
            & \multicolumn{1}{|l|}{+AdaptGuard} & 92.8 & 84.0 & 72.6 & 94.7 & 73.5 & 97.0 & 85.8 \\
            \midrule
            \multirow{5}{5.0em}{\textbf{Blended} \cite{chen2017targeted} Attack (SHOT)} & \multicolumn{1}{|l|}{SHOT \cite{liang2020we}} & 1.7 & 4.1 & 38.8 & 16.6 & 36.3 & 3.5 & 16.8 \\
            & \multicolumn{1}{|l|}{+ANP \cite{wu2021adversarial}} & 3.9 & 9.1 & 44.1 & 29.0 & 18.9 & 3.7 & 18.1 \\
            & \multicolumn{1}{|l|}{+TRADES \cite{zhang2019theoretically}} & 16.5 & 30.6 & 50.4 & 50.1 & 47.1 & 15.2 & 35.0 \\
            & \multicolumn{1}{|l|}{+PGD \cite{madry2017towards}} & 18.7 & 32.3 & 54.0 & 44.4 & 52.9 & 20.6 & 37.1 \\
            & \multicolumn{1}{|l|}{+AdaptGuard} & \best{59.1} & \best{61.6} & \best{56.8} & \best{69.5} & \best{57.9} & \best{64.2} & \best{61.5} \\

            \midrule
            \midrule
          
            \multirow{5}{5.0em}{Clean (NRC)} & \multicolumn{1}{|l|}{NRC \cite{yang2021exploiting}} & 91.8 & 90.8 & \bestclean{75.7} & \bestclean{97.9} & \bestclean{74.6} & 99.0 & \bestclean{88.3} \\
            & \multicolumn{1}{|l|}{+ANP \cite{wu2021adversarial}} & 90.0 & 92.2 & 75.3 & 97.1 & 73.8 & 98.8 & 87.9 \\
            & \multicolumn{1}{|l|}{+TRADES \cite{zhang2019theoretically}} & 90.6 & 88.4 & 72.5 & 97.4 & 68.1 & 99.4 & 86.1 \\
            & \multicolumn{1}{|l|}{+PGD \cite{madry2017towards}} & 93.0 & \bestclean{91.5} & 68.3 & 96.5 & 69.4 & \bestclean{99.6} & 86.4 \\
            & \multicolumn{1}{|l|}{+AdaptGuard} & \bestclean{95.0} & 87.0 & 73.8 & 96.0 & 73.4 & 98.2 & 87.2 \\
            \midrule
            \multirow{5}{5.0em}{\textbf{Blended} \cite{chen2017targeted} Attack (NRC)} & \multicolumn{1}{|l|}{NRC \cite{yang2021exploiting}} & 1.2 & 1.3 & 31.1 & 7.7 & 27.1 & 2.7 & 11.8 \\
            & \multicolumn{1}{|l|}{+ANP \cite{wu2021adversarial}} & 2.3 & 3.5 & 33.7 & 13.5 & 18.5 & 2.1 & 12.2 \\
            & \multicolumn{1}{|l|}{+TRADES \cite{zhang2019theoretically}} & 22.6 & 33.7 & 47.6 & 46.1 & 45.3 & 18.5 & 35.6 \\
            & \multicolumn{1}{|l|}{+PGD \cite{madry2017towards}} &  26.3 & 36.7 & \best{54.5} & 50.1 & 52.4 & 32.7 & 42.1 \\
            & \multicolumn{1}{|l|}{+AdaptGuard} & \best{56.2} & \best{61.2} & 53.3 & \best{69.3} & \best{54.1} & \best{64.8} & \best{59.8} \\
            
            \bottomrule
            \end{tabular}
        }
    \label{tab: result office Blended} 
    \end{table*}

\setlength{\tabcolsep}{3.0pt}
    \begin{table*}[!t]
        \centering
        \caption{Accuracies (\%) of defense methods against \textbf{SIG} \cite{barni2019new} backdoor attack on \textbf{OfficeHome} \cite{venkateswara2017deep} for model adaptation (ResNet-50).}
        \vspace{3mm}
        \resizebox{0.95\textwidth}{!}{
            \begin{tabular}{ll|cccccccccccca}
            \toprule
            \multicolumn{2}{c|}{Task} & A$\to$C & A$\to$P & A$\to$R & C$\to$A & C$\to$P & C$\to$R & P$\to$A & P$\to$C & P$\to$R & R$\to$A & R$\to$C & R$\to$P & Avg \\ 
            \midrule
            \multicolumn{2}{c|}{Clean (Source Only)} & 43.5 & 63.7 & 72.0 & 48.3 & 59.1 & 61.9 & 50.0 & 39.1 & 71.8 & 62.7 & 44.0 & 75.6 & 57.6 \\
            \multicolumn{2}{c|}{\textbf{SIG} \cite{barni2019new} Attack (Source Only)} & 1.7 & 2.5 & 6.0 & 5.7 & 0.9 & 6.1 & 4.6 & 1.1 & 5.1 & 2.2 & 0.5 & 1.1 & 3.1\\
            \midrule
            \midrule        
            \multirow{5}{3.5em}{Clean (SHOT)} & \multicolumn{1}{|l|}{SHOT \cite{liang2020we}} &  \bestclean{56.1} & \bestclean{77.7} & \bestclean{80.6} & \bestclean{66.8} & \bestclean{76.0} & \bestclean{78.2} & \bestclean{67.5} & 53.4 & \bestclean{82.4} & \bestclean{73.6} & \bestclean{58.0} & \bestclean{83.6} & \bestclean{71.2} \\
            & \multicolumn{1}{|l|}{+ANP \cite{wu2021adversarial}} & 53.7 & 75.7 & 78.0 & 60.7 & 63.6 & 70.9 & 65.6 & \bestclean{53.8} & 81.0 & 3.5 & 2.1 & 1.6 & 50.8 \\
            & \multicolumn{1}{|l|}{+TRADES \cite{zhang2019theoretically}} & \bestclean{56.1} & 75.8 & 79.1 & \bestclean{66.8} & 74.9 & 76.9 & 67.2 & 53.2 & 81.1 & 73.1 & 57.0 & 81.3 & 70.2 \\
            & \multicolumn{1}{|l|}{+PGD \cite{madry2017towards}} & 54.5 & 74.0 & 77.2 & 66.7 & 72.7 & 74.6 & 66.8 & 51.8 & 79.4 & 73.4 & 55.6 & 80.6 & 68.9 \\
            & \multicolumn{1}{|l|}{+AdaptGuard} & 54.7 & 75.3 & 78.8 & 63.1 & 72.9 & 73.9 & 63.3 & 49.7 & 79.4 & 68.2 & 55.7 & 80.5 & 68.0 \\
            \midrule
       
            \multirow{5}{3.5em}{\textbf{SIG} \cite{barni2019new} Attack (SHOT)} & \multicolumn{1}{|l|}{SHOT \cite{liang2020we}} & 5.0 & 39.3 & 52.6 & 46.5 & 15.2 & 48.0 & 48.8 & 5.6 & 50.8 & 43.1 & 2.3 & 30.7 & 32.3 \\
            & \multicolumn{1}{|l|}{+ANP \cite{wu2021adversarial}} & 36.9 & 54.5 & 61.4 & 41.6 & 46.1 & 53.2 & 45.1 & 29.9 & 62.1 & 1.1 & 2.2 & 0.8 & 36.2  \\
            & \multicolumn{1}{|l|}{+TRADES \cite{zhang2019theoretically}} & 45.6 & 58.3 & 65.8 & 51.6 & 58.3 & 59.6 & 52.9 & 45.7 & 68.4 & 55.2 & 46.5 & 59.8 & 55.6 \\
            & \multicolumn{1}{|l|}{+PGD \cite{madry2017towards}} & \best{49.8} & 66.4 & 70.1 & 58.4 & 66.0 & \best{70.1} & 60.0 & \best{50.2} & 73.8 & 59.3 & \best{50.5} & 70.9 & 62.1  \\
            & \multicolumn{1}{|l|}{+AdaptGuard} & 44.6 & \best{67.3} & \best{75.0} & \best{61.1} & \best{66.3} & 70.0 & \best{61.2} & 41.5 & \best{76.2} & \best{65.9} & 47.1 & \best{72.2} & \best{62.4} \\
            \midrule
            \midrule        
            \multirow{5}{3.5em}{Clean (NRC)} & \multicolumn{1}{|l|}{NRC \cite{yang2021exploiting}} & \bestclean{56.4} & \bestclean{78.5} & \bestclean{80.2} & \bestclean{66.4} & \bestclean{77.4} & \bestclean{78.6} & \bestclean{65.1} & \bestclean{57.3} & \bestclean{80.7} & \bestclean{72.2} & \bestclean{57.8} & \bestclean{83.2} & \bestclean{71.1} \\
            & \multicolumn{1}{|l|}{+ANP \cite{wu2021adversarial}} & 56.0 & 77.5 & 78.9 & 61.9 & 75.0 & 74.4 & 62.7 & 57.0 & 80.2 & 4.7 & 4.2 & 5.0 & 53.1  \\
            & \multicolumn{1}{|l|}{+TRADES \cite{zhang2019theoretically}} & 49.6 & 69.8 & 70.2 & 56.9 & 69.7 & 65.8 & 51.6 & 46.6 & 71.7 & 64.6 & 52.7 & 76.3 & 62.1 \\
            & \multicolumn{1}{|l|}{+PGD \cite{madry2017towards}} &  50.3 & 70.4 & 70.9 & 64.9 & 71.6 & 69.2 & 61.9 & 48.8 & 75.3 & 68.4 & 50.9 & 76.6 & 64.9 \\
            & \multicolumn{1}{|l|}{+AdaptGuard} & 55.5 & 75.3 & 78.6 & 60.7 & 72.5 & 72.1 & 61.5 & 50.5 & 77.0 & 66.1 & 56.8 & 80.2 & 67.2 \\
            \midrule
        
            \multirow{5}{3.5em}{\textbf{SIG} \cite{barni2019new} Attack (NRC)} & \multicolumn{1}{|l|}{NRC \cite{yang2021exploiting}} &   2.6 & 27.3 & 33.8 & 42.2 & 20.2 & 32.8 & 37.1 & 4.8 & 32.9 & 33.4 & 6.9 & 16.9 & 24.2 \\
            & \multicolumn{1}{|l|}{+ANP \cite{wu2021adversarial}} & 28.2 & 40.9 & 46.4 & 44.1 & 30.3 & 41.5 & 36.6 & 11.1 & 33.7 & 2.7 & 3.7 & 3.9 & 26.9 \\
            & \multicolumn{1}{|l|}{+TRADES \cite{zhang2019theoretically}} & 25.1 & 33.6 & 44.1 & 27.5 & 34.1 & 37.0 & 22.9 & 25.8 & 39.4 & 28.8 & 20.9 & 36.6 & 31.3 \\
            & \multicolumn{1}{|l|}{+PGD \cite{madry2017towards}} & \best{44.0} & 55.8 & 52.3 & 46.3 & 61.9 & 56.4 & 38.0 & \best{43.2} & 57.5 & 37.5 & 39.1 & 58.2 & 49.2 \\
            & \multicolumn{1}{|l|}{+AdaptGuard} & 42.0 & \best{68.1} & \best{73.3} & \best{56.9} & \best{65.6} & \best{67.1} & \best{57.1} & 38.7 & \best{72.0} & \best{63.2} & \best{49.4} & \best{73.7} & \best{60.6} \\
            
            \bottomrule
            \end{tabular}
        }
    \label{tab: result office-home SIG} 
    \end{table*}

\setlength{\tabcolsep}{4.0pt}
    \begin{table*}[!t]
        \centering
        \caption{Accuracies (\%) of defense methods against \textbf{SIG} \cite{barni2019new} backdoor attack on \textbf{Office} \cite{saenko2010adapting} dataset for model adaptation (ResNet-50).}
        \vspace{3mm}
        \resizebox{0.7\textwidth}{!}{
            \begin{tabular}{ll|cccccca}
            \toprule
            \multicolumn{2}{c|}{Task} & A$\to$D & A$\to$W & D$\to$A & D$\to$W & W$\to$A & W$\to$D & Avg \\ 
            \midrule       
            \multicolumn{2}{c|}{Clean (Source Only)} & 79.7 & 74.8 & 60.3 & 93.8 & 61.5 & 96.8 & 77.8 \\
            \multicolumn{2}{c|}{\textbf{SIG} \cite{barni2019new} Attack (Source Only)} &  0.8 & 1.4 & 22.3 & 4.7 & 5.4 & 2.5 & 6.2   \\

            \midrule
            \midrule
         
            \multirow{5}{3.5em}{Clean (SHOT)} & \multicolumn{1}{|l|}{SHOT \cite{liang2020we}} & \bestclean{95.0} & \bestclean{92.3} & \bestclean{74.3} & 96.5 & \bestclean{75.0} & 99.0 & \bestclean{88.7} \\
            & \multicolumn{1}{|l|}{+ANP \cite{wu2021adversarial}} & 93.4 & 84.7 & 74.1 & 95.9 & 74.8 & \bestclean{99.6} & 87.0 \\
            & \multicolumn{1}{|l|}{+TRADES \cite{zhang2019theoretically}} & 93.0 & 90.8 & 73.9 & 95.5 & 72.7 & 97.2 & 87.2 \\
            & \multicolumn{1}{|l|}{+PGD \cite{madry2017towards}} & 94.8 & \bestclean{92.3} & 72.2 & \bestclean{96.9} & 72.0 & 99.0 & 87.9 \\
            & \multicolumn{1}{|l|}{+AdaptGuard} & 92.8 & 83.4 & 72.7 & 93.3 & 72.1 & 97.2 & 85.2 \\
            \midrule
        
            \multirow{5}{3.5em}{\textbf{SIG} \cite{barni2019new} Attack (SHOT)} & \multicolumn{1}{|l|}{SHOT \cite{liang2020we}} & 9.9 & 20.5 & 67.2 & 54.1 & 56.1 & 10.9 & 36.4 \\
            & \multicolumn{1}{|l|}{+ANP \cite{wu2021adversarial}} & 8.9 & 20.8 & 62.9 & 58.0 & 58.6 & 13.6 & 37.1 \\
            & \multicolumn{1}{|l|}{+TRADES \cite{zhang2019theoretically}} & 27.4 & 50.5 & 69.3 & 80.7 & 65.9 & 29.6 & 53.9 \\
            & \multicolumn{1}{|l|}{+PGD \cite{madry2017towards}} & 39.5 & \best{62.1} & \best{69.6} & \best{86.0} & \best{68.5} & 37.0 & \best{60.5} \\
            & \multicolumn{1}{|l|}{+AdaptGuard} & \best{57.8} & 61.6 & 56.7 & 66.2 & 56.3 & \best{64.4} & \best{60.5} \\

            \midrule
            \midrule
   
            \multirow{5}{3.5em}{Clean (NRC)} & \multicolumn{1}{|l|}{NRC \cite{yang2021exploiting}} & 89.8 & 90.4 & \bestclean{76.0} & \bestclean{97.5} & \bestclean{74.8} & 99.0 & \bestclean{87.9} \\
            & \multicolumn{1}{|l|}{+ANP \cite{wu2021adversarial}} & 90.4 & 89.3 & 75.6 & \bestclean{97.5} & \bestclean{74.8} & \bestclean{99.8} & \bestclean{87.9} \\
            & \multicolumn{1}{|l|}{+TRADES \cite{zhang2019theoretically}} & 87.8 & 87.6 & 70.5 & 97.0 & 67.3 & 98.0 & 84.7 \\
            & \multicolumn{1}{|l|}{+PGD \cite{madry2017towards}} & 92.8 & \bestclean{91.3} & 66.5 & 96.5 & 68.6 & 99.8 & 85.9 \\
            & \multicolumn{1}{|l|}{+AdaptGuard} & \bestclean{93.2} & 86.0 & 72.4 & 93.0 & 72.6 & 98.0 & 85.9 \\
            \midrule
            
            \multirow{5}{3.5em}{\textbf{SIG} \cite{barni2019new} Attack (NRC)} & \multicolumn{1}{|l|}{NRC \cite{yang2021exploiting}} &  4.7 & 14.0 & 52.8 & 30.8 & 35.6 & 4.7 & 23.8 \\
            & \multicolumn{1}{|l|}{+ANP \cite{wu2021adversarial}} & 6.2 & 19.5 & 54.8 & 38.5 & 44.3 & 8.0 & 28.5 \\
            & \multicolumn{1}{|l|}{+TRADES \cite{zhang2019theoretically}} & 30.0 & 56.0 & 63.9 & 82.4 & 54.4 & 38.9 & 54.3 \\
            & \multicolumn{1}{|l|}{+PGD \cite{madry2017towards}} & 49.0 & 67.0 & 64.7 & 87.7 & 66.7 & 57.6 & 65.4  \\
            & \multicolumn{1}{|l|}{+AdaptGuard} & \best{86.0} & \best{82.4} & \best{67.0} & \best{90.3} & \best{68.3} & \best{95.1} & \best{81.5} \\
            
            \bottomrule
            \end{tabular}
        }
    \label{tab: result office SIG} 
    \end{table*}

\end{document}